\documentclass[reprint,
superscriptaddress,
 bibnotes,
 amsmath,amssymb,
 aps,
 prapplied,
longbibliography 
]{revtex4-2}

\usepackage{graphics,graphicx,float,amsmath, xcolor,upgreek,hyperref,xr-hyper,bm}
\usepackage{gensymb}

\begin{document}

\title{Cavity-Enhanced Photon Emission from a Single Germanium-Vacancy Center in a Diamond Membrane}

\author{Rasmus H$\o$y Jensen}
\thanks{R. H. J. and E. J. contributed equally to this work.}
\email[corresponding authors: ]{rasjen@fysik.dtu.dk and erika.janitz@mail.mcgill.ca}
\affiliation{Center for Macroscopic Quantum States (bigQ), Department of Physics, Technical University of Denmark, Lyngby, Denmark}

\author{Erika Janitz}
\thanks{R. H. J. and E. J. contributed equally to this work.}
\email[corresponding authors: ]{rasjen@fysik.dtu.dk and erika.janitz@mail.mcgill.ca}
\affiliation{Department of Physics, McGill University, Montreal, Quebec, Canada}

\author{Yannik Fontana}
\affiliation{Center for Macroscopic Quantum States (bigQ), Department of Physics, Technical University of Denmark, Lyngby, Denmark}

\author{Yi He}
\affiliation{Department of Electrical and Computer Engineering, Carnegie Mellon University, Pittsburgh, Pennsylvania 15213, USA}

\author{Olivier Gobron}
\affiliation{Center for Macroscopic Quantum States (bigQ), Department of Physics, Technical University of Denmark, Lyngby, Denmark}

\author{Ilya P. Radko}
\affiliation{Center for Macroscopic Quantum States (bigQ), Department of Physics, Technical University of Denmark, Lyngby, Denmark}

\author{Mihir Bhaskar}
\affiliation{Department of Physics, Harvard University, Cambridge, Massachusetts 02138, USA}

\author{Ruffin Evans}
\affiliation{Department of Physics, Harvard University, Cambridge, Massachusetts 02138, USA}

\author{C$\text{\'e}$sar Daniel Rodr$\text{\'i}$guez Rosenblueth}
\affiliation{Department of Physics, McGill University, Montreal, Quebec, Canada}

\author{Lilian Childress}
\affiliation{Department of Physics, McGill University, Montreal, Quebec, Canada}

\author{Alexander Huck}
\affiliation{Center for Macroscopic Quantum States (bigQ), Department of Physics, Technical University of Denmark, Lyngby, Denmark}

\author{Ulrik Lund Andersen}
\affiliation{Center for Macroscopic Quantum States (bigQ), Department of Physics, Technical University of Denmark, Lyngby, Denmark}


\date{\today}

\begin{abstract}
The nitrogen-vacancy center in diamond has been explored extensively as a light-matter interface for quantum information applications, however it is limited by low coherent photon emission and spectral instability. Here, we present a promising interface based on an alternate defect with superior optical properties (the germanium-vacancy) coupled to a finesse $\approx 11{,}000$ fiber cavity, resulting in a $31^{+11}_{-15}$\,-fold increase in the spectral density of zero phonon line emission. This work sets the stage for cryogenic experiments, where we predict a measurable increase in the spontaneous emission rate.

\end{abstract}

\maketitle
Defect centers in diamond can exhibit long-lived spin states that are accessible via coherent optical transitions \cite{doherty2013,rose2018,sukachev2017}, providing a promising platform for quantum nonlinear optics \cite{chang2014} or quantum networks \cite{kimble2008,wehner2018}. There have been impressive steps towards creating such a network using the nitrogen-vacancy (NV) center in diamond, with several critical components demonstrated in the last decade \cite{togan2010,bernien2012,bernien2013,hensen2015,pfaff2014,maurer2012,kalb2017}. Nevertheless, optically-mediated entanglement rates for NVs have been limited to tens of Hz \cite{humphreys2018} due to the low fraction ($3\%$) of photons emitted into the coherent zero phonon line (ZPL), the difficulty of collecting these photons into a single optical mode, the long excited state lifetime ($12$ ns), and spectral diffusion of the emitter. This platform could be improved by coupling defects to an optical resonator, thereby increasing the ZPL emission rate into a well-defined mode. However, enhanced spectral diffusion near surfaces has thus far impeded attempts to realize stable NV centers in small-mode-volume optical cavities \cite{faraon2012,li2015,riedel2017}. 

An alternative approach explores group-IV defect centers such as the silicon-vacancy (SiV) \cite{hepp2014,rogers2014} and germanium-vacancy (GeV) centers \cite{siyushev2017,iwasaki2015}. While their spins interact strongly with phonons (necessitating low temperature \cite{sukachev2017} or high-strain \cite{sohn2018}), their optical properties are superior. These defects exhibit larger ZPL fractions ($60-70$\%) \cite{Thiering2017}, and strongly reduced spectral diffusion owing to their inversion symmetry \cite{evans2016,Bhaskar2017,wan2019}. Moreover, there is evidence that the GeV also has high quantum efficiency \cite{Bhaskar2017}. Integrating GeVs into optical resonators is therefore a promising step toward generating efficient or even deterministic spin-photon interactions \cite{evans2018}.
     
Recently, two primary approaches for coupling diamond defects to optical cavities have emerged: nanophotonics and open-geometry microcavities. Emitters coupled to low-mode-volume nanophotonic resonators can exhibit high Purcell enhancement \cite{faraon2012} and can even enter the high-cooperativity regime \cite{sipahigil2016,evans2018}. Despite their inversion symmetry, SiVs in such structures still suffer from increased spectral diffusion and inhomogeneous broadening due to nearby surfaces, lattice damage, and material strain \cite{nguyen2019}. This is further exacerbated for the GeV, since the heavier ion leads to increased implantation damage. Furthermore, the shorter ZPL wavelength ($603$ nm) makes fabrication more challenging, requiring smaller feature sizes with increased sensitivity to surface roughness. 
\begin{figure}[ht!]
\includegraphics{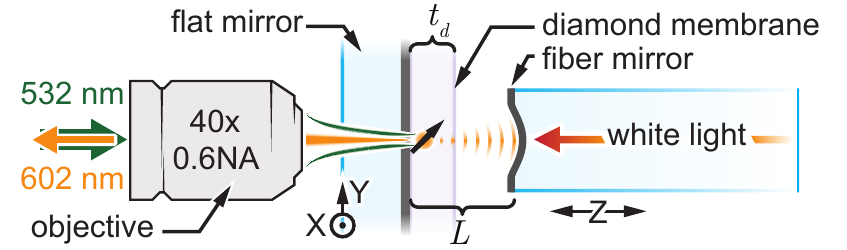}
\caption{A schematic of the cavity setup, where $L$ is the cavity length and $t_d=862^{+1}_{-4}$ nm is the membrane thickness.\label{setup}}
\end{figure}

In contrast to nanofabricated devices, open Fabry-P{\'e}rot microcavities offer narrower resonator linewidths and in-situ spectral and spatial tunability. Importantly, these systems accommodate microns-thick diamond membranes in which even NVs can exhibit minimal spectral diffusion and bulk-like optical properties \cite{ruf2019}. Thus far, it has proven challenging to fabricate membranes with sufficient surface quality to achieve very high cavity finesse. This is particularly true when the optical energy is concentrated in the diamond (so-called ``diamond-like" modes) due to a field antinode at the air-diamond interface \cite{janitz2015}. Approaches using thinner (sub-wavelength) membranes may relax constraints on surface quality by precluding ``diamond-like" mode confinement \cite{haussler2019}, but exposes defects to increased surface noise and material stress. Indeed, the only demonstration of open cavity-coupling of a single emitter in a diamond membrane used a finesse $\mathcal{F}=5{,}260$ mode localized in air \cite{riedel2017}. Already, this system achieved a $30$-fold enhancement in NV center ZPL emission and a factor of two reduction in the excited state lifetime, illustrating the potential of the open-cavity approach.

In this work, we observe coupling between a ``diamond-like" mode of a Fabry-P\'{e}rot microcavity and a single GeV center in a membrane, thereby combining the advantageous optical properties of the GeV with the flexibility of an open-geometry resonator. We build on previous room-temperature experiments using nanodiamonds \cite{Albrecht2013,benedikter2017} to observe funneling of emission into the cavity mode. Our experiment represents the first demonstration of single GeV-cavity coupling and confirms that both the diamond membrane and cavity are of sufficient quality to support high-finesse ($\mathcal{F}>$10{,}000) ``diamond-like" resonances. Furthermore, we observe and study the presence of a dark state in the GeV level structure, elucidating new details regarding the emission dynamics. These results therefore represent an important step toward realizing an efficient spin-photon interface using diamond defects in open cavities.

Our microcavities comprise a macroscopic flat mirror and a microscopic spherical mirror fabricated on the tip of a single-mode optical fiber (Thorlabs 630HP). The spherical dimple of the fiber mirror is machined using a laser ablation technique \cite{hunger2010} with an effective radius of curvature $R=43.1\pm0.6\ \upmu$m \footnote{\label{suppnote}See Supplemental Material at [URL will be inserted by publisher] for further information regarding membrane fabrication, experimental characterization, and theoretical details. The supplementary includes Refs. [50-62]}. Both substrates are coated with dielectric Bragg stack mirrors (LASEROPTIK); the flat mirror is low-index terminated such that the addition of the diamond membrane increases the mirror reflectivity, yielding an electric field antinode at the mirror-diamond interface. In contrast, the fiber mirror is high-index terminated with a field node at the air-mirror interface. The layers are numerically optimized to have matched transmissions $T_{flat}=T_{fiber}=70$ ppm at 603 nm and $T_{flat}=0.998$ and $T_{fiber}= 0.007 $ at the excitation wavelength (532 nm).

We incorporate emitters into the cavity by Van der Waals bonding a diamond membrane containing GeVs to the flat mirror \footnotemark[\value{footnote}], which is mounted on a three-axis stage with piezoelectric control (see Fig. \ref{setup}). A long working distance objective (Olympus LUCPLFLN $40\times$, $0.6$ NA) faces the backside of the flat mirror and is used for emitter excitation and cavity mode collection; a separate stage facing the sample holds either an objective (Mitutoyo $100\times$ Plan Apo, $0.7$ NA) or the fiber mirror for confocal or cavity characterization respectively. Furthermore, we use a piezoelectric scanner along the optical axis for fine tuning the cavity length and align the objectives and fiber mirror to the excitation laser, scanning the sample relative to the excitation spot to study different emitters.
%

\begin{figure}
	\includegraphics{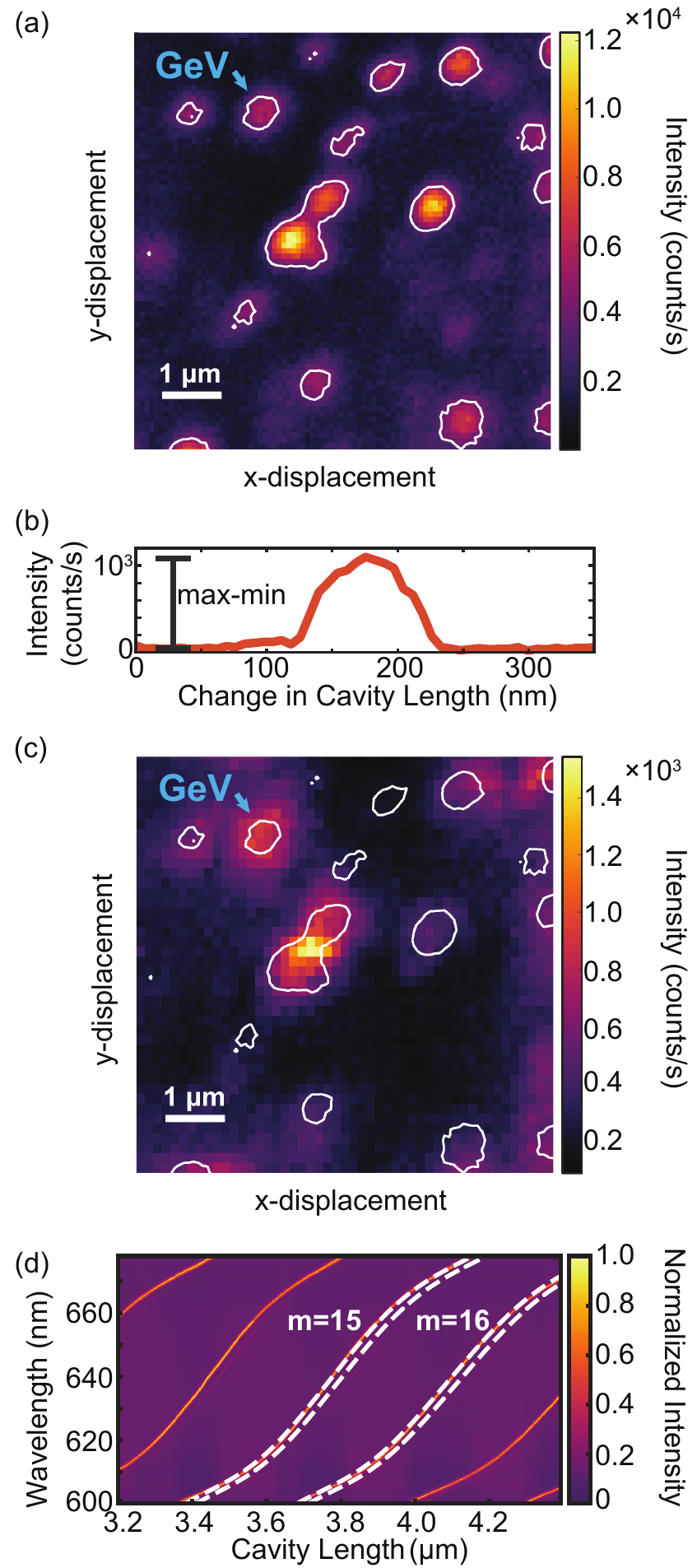}
	\caption{a) A confocal image of membrane fluorescence ($600-605$ nm, $P=19$ mW). The white contours illustrate emitter localization and the studied GeV is indicated with an arrow. b) Cavity-coupled emission as a function of resonator length for the studied defect ($\mathcal{F}=11{,}200\pm1{,}700$). The non-Lorentzian shape is determined by filters in the collection path. c) A map of cavity-coupled emission amplitude (max-min) over the same area as in a) ($P=19$ mW). The confocal contours are overlaid, and the studied GeV indicated. d) Broadband cavity transmission as a function of cavity length. The dashed lines correspond to numerical fits of the fundamental and first order transverse modes of the $m=15$ and $16$ longitudinal modes. \label{fig2}}  
\end{figure}

We first characterize the membrane in the confocal configuration to map out the position of GeVs. Confocal images are obtained by exciting emitters through the back of the flat mirror, while the emission is collected through the $100\times$ objective and filtered with a passband of $600-605$ nm before coupling into a single-mode fiber (Fig. \ref{fig2}a). Here, bright spots correspond to single GeVs and contours are drawn at $4{,}000$ counts/s to qualitatively show emitter localization. We perform a detailed study of the well-isolated GeV center marked with an arrow (henceforth referred to as the studied emitter). We confirmed that this was a single defect (measured intensity autocorrelation of $g^2(0)=0.25 \pm 0.16$ \footnotemark[\value{footnote}]) with a lifetime of $\tau=6.0\pm0.1$ ns. 

\begin{figure}[h]
	\centering
	\includegraphics{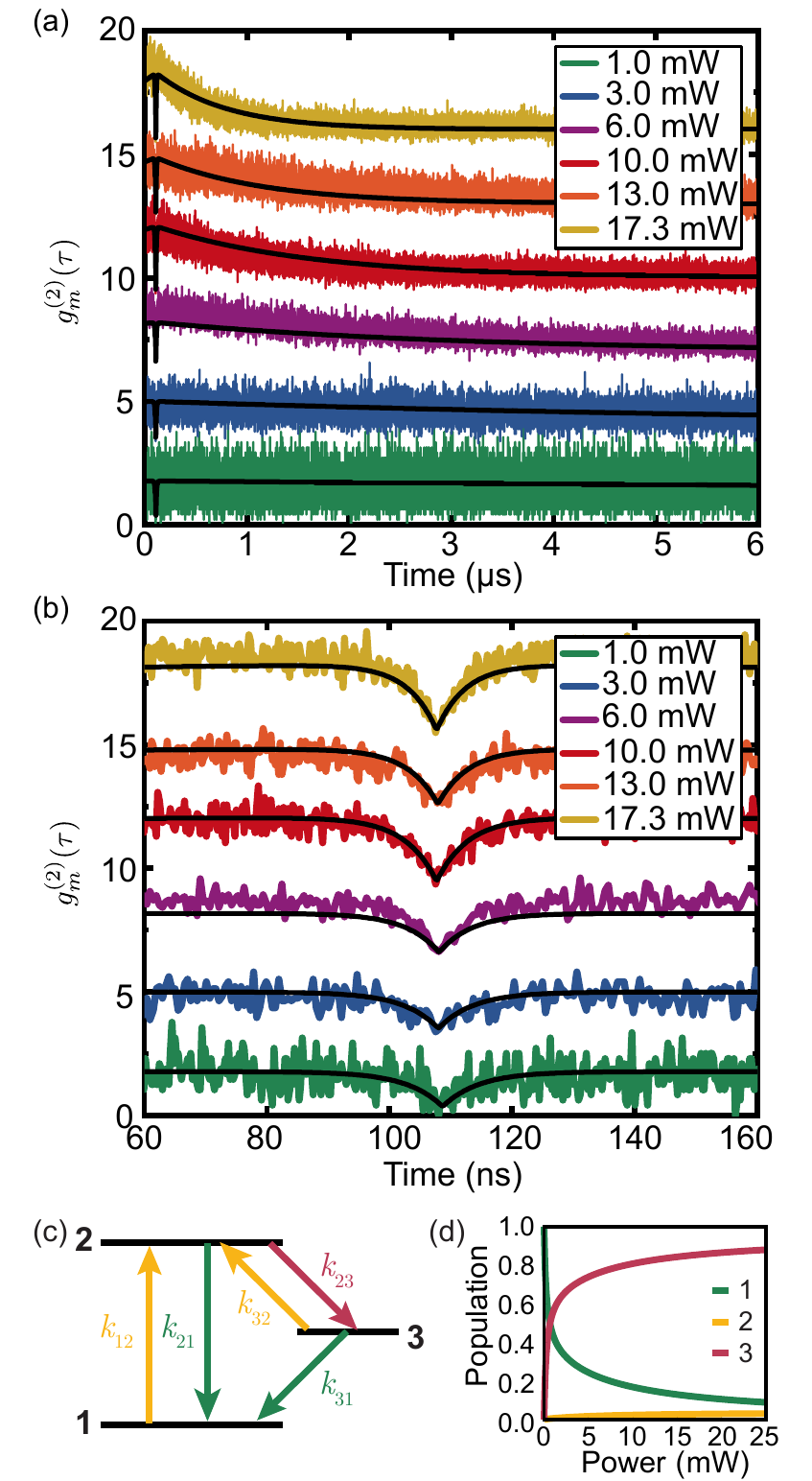}
	\caption{a) Power dependent $g_m^{(2)}$ for the studied GeV (subsequent measurements are offset by 3 for clarity). Fits to Eq. \ref{eq:g2meas} are shown in black. b) $g_m^{(2)}$ data and fits plotted at shorter time-scales. c) Equivalent three-level system for rate analysis. d) State populations as a function of pump power. 
		\label{fig3}}
\end{figure}

The modest fluorescence count rates observed can be attributed to a combination of low transmission in the collection path, narrow spectral filtering, and much of the emission either exiting through the flat mirror (which is designed for normal incidence) or being confined to the membrane in the form of guided modes \footnotemark[\value{footnote}].
Counts are further reduced by the presence of an additional dark state, which may be a different charge state of the GeV center \cite{chen2019}. To quantify the impact of such a state, we probed the population dynamics of the studied emitter by measuring the second order correlation function for varying excitation power (shown in Figs. \ref{fig3}a and b for different timescales). We fit this data assuming a Poissonian background contribution parameterized by $\sigma=S/(S+B)$, where $ S $ is the signal from the GeV center and $ B $ is background \cite{Brouri2000}, resulting in
\begin{align}
g_m^{(2)}(\tau) &= g^{(2)}(\tau)\, \sigma^2 + 1-\sigma^2,\label{eq:g2meas}
\end{align}
where
\begin{align}
g^{(2)}(\tau) &= 1 - (1+a)\,e^{-|\tau|/\tau_1}  + a\, e^{-|\tau|/\tau_2}.
\label{eq:g2fit}
\end{align}
The bunching of the $g^{(2)}_m$ function at high excitation powers implies that there are three or more states participating in the dynamics; we found that the system was well-described by a three-level model including power-dependent shelving (Fig. \ref{fig3}c, further details in \footnotemark[\value{footnote}]) \cite{Neu2012,benedikter2017}.

From this analysis, we calculate the populations as a function of excitation power (Fig. \ref{fig3}d). We predict a dark-state equilibrium population of $96\pm18\%$ at infinite $532$ nm pump power, corresponding to a bright-state photon emission rate of $(6.8\pm0.9)\times 10^6$ photons/s. Reassuringly, there has been some evidence that the dark-state population can be reduced via laser repumping/gating \cite{chen2019,sipahigil2016}. Moreover, non-radiative decay paths may further limit photon counts, as varying estimates of quantum efficiency have been reported recently \cite{Bhaskar2017,chen2019,boldyrev2018, Nguyen2019-gev}. Finally, photon emission was further limited by our inability to saturate most emitters due to the low absorption cross-section at $532$ nm \cite{hausler2017}.
 
 We then replace the $100\times$ objective with the fiber mirror and measure a cavity finesse of $\mathcal{F}=11{,}200\pm1{,}700$ at $603$ nm \cite{janitz2015}. Exciting the emitter with $532$ nm, we observe cavity-coupled emission as a function of cavity length (Fig. \ref{fig2}b). The counts clearly peak when the cavity length is resonant with the thermally broadened ZPL; optical filtering results in a profile that deviates from the expected Lorentzian. We repeat this measurement at various positions on the membrane and plot the difference in measured counts in Fig. \ref{fig2}c, where the lateral resolution is set by the excitation spot size ($\approx1$ $\upmu$m) since the cavity waist diameter is larger ($\approx3$ $\upmu$m). The contours obtained from the confocal measurement are overlaid for comparison, revealing a clear spatial correspondence between emitters in the confocal scan and high emission into the cavity. This measurement corresponds to emission into the $m = 15$ longitudinal cavity mode, as determined by numerical fitting of the broadband cavity transmission measurements (Fig. \ref{fig2}d) \footnotemark[\value{footnote}].   

\begin{figure}
\includegraphics{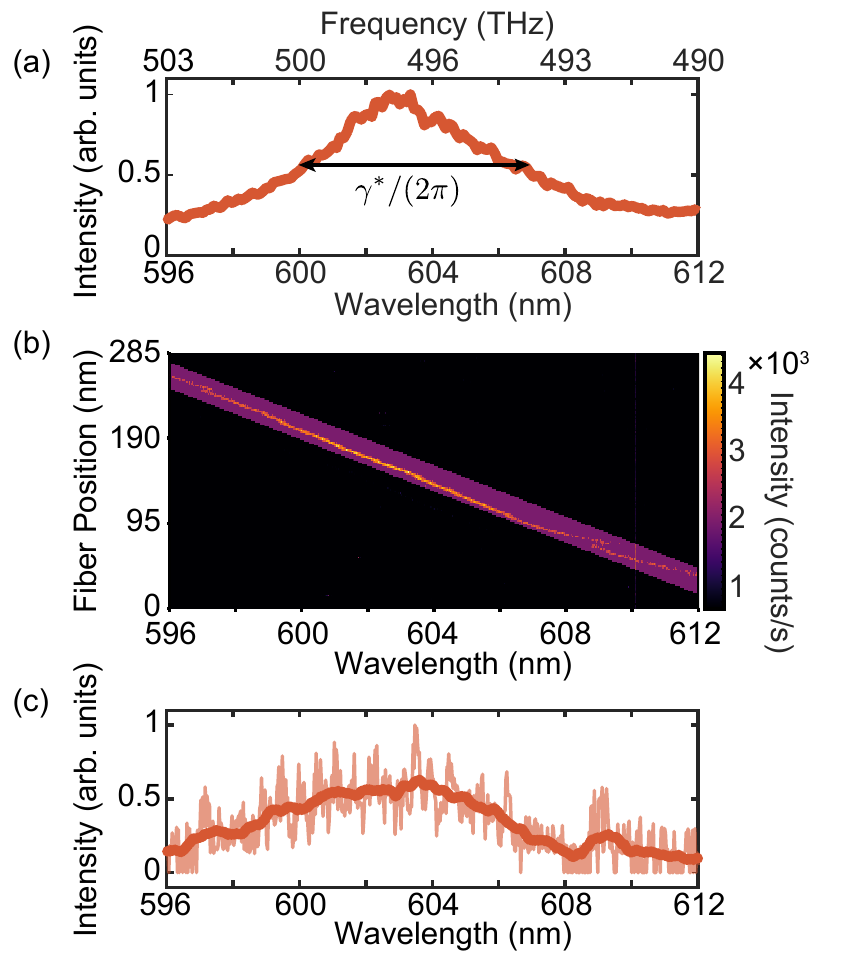}
\caption{a) GeV center spectrum taken in the confocal configuration. The pure dephasing rate $\gamma^*/(2\pi)= 5.22\pm0.05$ THz is indicated. b) GeV center fluorescence as a function of cavity length. The colored region represents the integration window. c) Integrated spectrum taken in the cavity configuration. The light trace shows the raw data, and the dark trace has been averaged over $40$ points.  \label{spectrum}}
\end{figure}

Further evidence that we are probing a cavity-coupled GeV can be obtained by comparing emitter spectra from both setups. Figure \ref{spectrum}a shows a confocal spectrum for the studied emitter, which exhibits a strong ZPL emission around $603$ nm with a FWHM linewidth of $\gamma^*/(2\pi)=5.22\pm0.05$ THz \footnotemark[\value{footnote}]. A cavity spectrum can be obtained by exciting the emitter while scanning the cavity length and acquiring a spectrum at each position (Fig. \ref{spectrum}b). Integrating the emission along one mode using a $2$ nm window about the resonance results in the light trace shown in Fig. \ref{spectrum}c, where the noise can be attributed to mechanical instability of the cavity; we also average the data using a $40$ point window to show the underlying shape (dark trace). The peaks around $597$ and $609$ nm are due to mechanical disturbances causing the cavity to pass through the same resonance twice. The qualitative similarities between the confocal and cavity spectra further confirm that the collected emission is coming from a GeV.

Comparison of cavity-coupled and free-space fluorescence rates allows us to quantify cavity funneling, however we must carefully account for different excitation intensities arising from interference effects in the diamond membrane and cavity. We thus compare the saturating fluorescence counts at infinite pump power ($I^{\infty}$) obtained by fitting saturation curves with the model $I(P) = I^{\infty} \tfrac{P}{P+P_{sat}} + c_{bg} P$, where $I$ is the observed count rate, $P$ is the excitation power, $P_{sat}$ is the saturation power, and $c_{bg}$ accounts for a linear background. In the confocal setup, we measure a saturation power of $P_{sat}=3.9 \pm 0.3$ mW and extract a saturating fluorescence count rate of $I^{\infty}_{free,meas}=4{,}000 \pm 200$ counts/s. We can calculate the total emission rate of the defect using our estimate of the collection efficiency for the confocal path ($\eta_{free}= (3.5^{+0.9}_{-1.5})\times 10^{-3}$ counts/photon) as $I^{\infty}_{free}=I^{\infty}_{free,meas}/\eta_{free}=(1.2^{+0.5}_{-0.3})\times 10^6$ photons/s  \footnotemark[\value{footnote}]. Comparing this value to the maximum photon emission rate predicted at infinite pump power results in a quantum efficiency of $17^{+8}_{-5}\%$ for the studied GeV, which falls within the range of reported values \cite{Bhaskar2017,chen2019,boldyrev2018}. Furthermore, we estimate the peak photon spectral density to be $90^{+40}_{-20}$ photons/(s GHz) \footnotemark[\value{footnote}].
\begin{figure}
\includegraphics{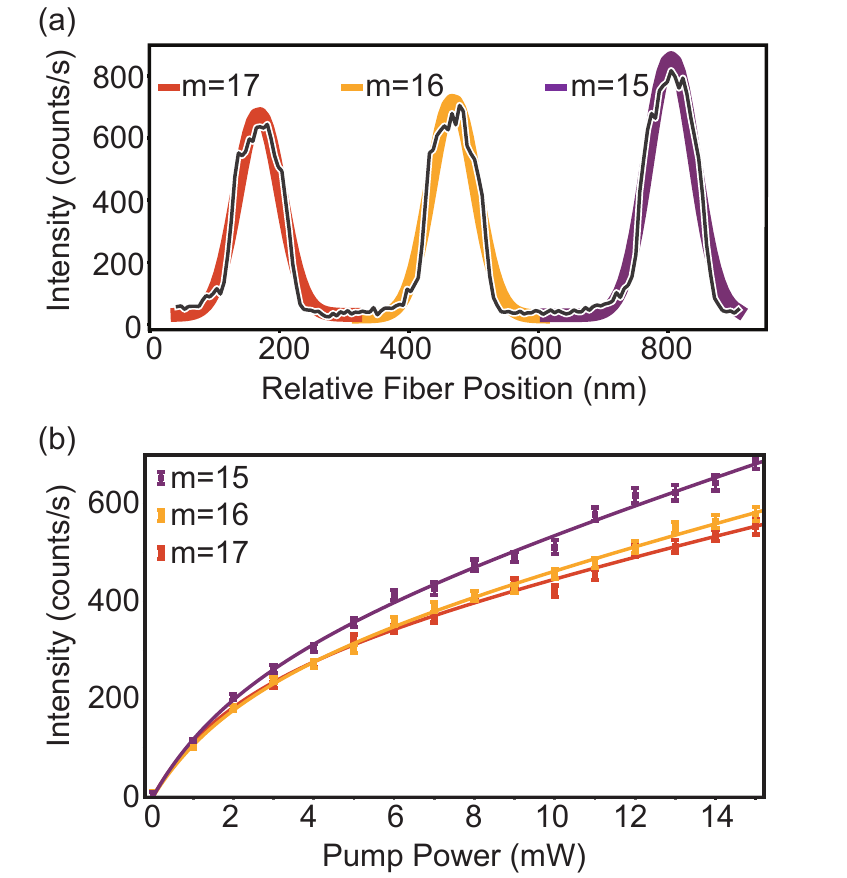}
\caption{a) A longitudinal cavity scan ($P=14$ mW). The amplitudes of the Gaussian fits for $m=15$, $16$, and $17$ are used to generate saturation curves. b) Corresponding saturation curves, where error bars indicate one standard deviation on the fit amplitude. \label{sat_fig}}
\end{figure}

We then study the saturation behavior of the emitter coupled to the $m=15$, $16$, and $17$ longitudinal modes of the cavity. For each excitation power, we measure fluorescence as a function of cavity length (Fig. \ref{sat_fig}a). The amplitudes of these resonances are estimated by fitting with Gaussians, and are plotted as a function of excitation power in Fig. \ref{sat_fig}b. As expected, the cavity-coupled fluorescence decreases with mode number as the cavity mode volume increases. Fitting the $m = 15$ data to the saturation model yields $P_{sat}=3.1 \pm 0.5$ mW and $I^{\infty}_{cav,meas}=380 \pm 50$ counts/s. We estimate a collection efficiency of $\eta_{cav}=(8.2 \pm 1.2)\times 10^{-2}$ counts/photon for the cavity setup, resulting in a corrected saturating fluorescence count rate of $I^{\infty}_{cav}=I^{\infty}_{cav,meas}/\eta_{cav}=4{,}700\pm900$ photons/s \footnotemark[\value{footnote}]. 
Moreover, we estimate a cavity linewidth of $\kappa/(2\pi)=1.08 \pm 0.17$ GHz, leading to a cavity-enhanced peak spectral density of $2{,}800 \pm 700$ photons/(s GHz), a factor of $31^{+11}_{-15}$ greater than what was obtained with the confocal measurements. The experimentally determined efficiency of emission into the $m=15$ mode is then $\beta_{exp}=I^{\infty}_{cav}/(I^{\infty}_{free}+I^{\infty}_{cav})=0.40^{+0.13}_{-0.19}\%$ \footnotemark[\value{footnote}].

\begin{figure}
    \centering
    \includegraphics{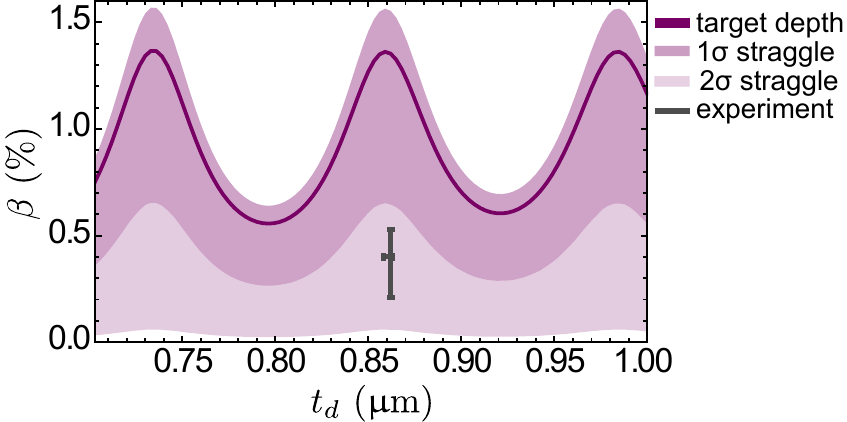}
    \caption{Simulated $\beta$ for an emitter resonantly coupled to the $m=15$ mode as a function of diamond thickness, with the experimentally derived value of $\beta_{exp}=0.40^{+0.13}_{-0.19}\%$ plotted at the extracted membrane thickness $t_d=862^{+1}_{-4}$ nm. }
    \label{sims}
\end{figure}

For comparison, we simulate $\beta$ for an emitter coupled to the $m=15$ cavity mode as a function of diamond thickness (Fig. \ref{sims}) \footnotemark[\value{footnote}]. The solid line shows the result for the target implantation depth ($125$ nm) and the shaded regions reveal the spread for one- and two-times the standard deviation in depth ($\sigma=20$ nm), resulting in $\beta$ between $0.06\%$ and $1.55\%$ at $t_d=862$ nm. The peaks in efficiency correspond to ``diamond-like" modes, where the field in the diamond is maximized. We note that the observed $\beta_{exp}$ lies at the lower range of simulated values, which we tentatively attribute to non-ideal emitter location along the cavity axis from implantation straggle.

Cooling such a high-quality membrane to cryogenic temperatures should result in a narrowing of the ZPL, approaching the radiative limit ($\gamma/(2\pi)=27$ MHz for $\tau=6.0$ ns) and allowing for a Purcell enhancement of cavity-coupled optical transitions. The expected enhancement can be calculated from the extracted room-temperature system parameters as $F_p\approx\frac{\gamma^*}{\xi\,\kappa}\beta_{exp}$, where $\xi=0.6$ is the GeV Debye-Waller factor \cite{palyanov2015}. This would result in $F_p=32^{+12}_{-16}$ (assuming unity quantum efficiency and full depopulation of the dark state), corresponding to a $20^{+7}_{-10}\,$-times reduction in the excited state lifetime and $>95\%$ of photons emitted into the ZPL. 
These values compare favorably with the current state-of-the-art for open microcavities, where the ZPL of an NV was enhanced by a comparable factor of $F_p=30$, with an excited state lifetime reduction of only two due to the lower branching ratio \cite{riedel2017}.

In summary, we demonstrate coupling of a single GeV center in a diamond membrane to a ``diamond-like" mode of an open-cavity system, resulting in an $31^{+11}_{-15}$-times increase in the spectral density of single-photon emission. Moving forward, stable cryogenic operation at the ZPL frequency should be achievable using readily available cavity-locking techniques \cite{janitz2017}; such resonant coupling would lead to a projected Purcell enhancement similar to what has been achieved for NV centers in open-cavities \cite{riedel2017} with a much larger reduction in the excited state lifetime. This could be further improved by reducing the cavity mode volume via the mirror radius of curvature, where a shallow mirror with a $R\approx5$ $\upmu$m would increase the enhancement by a factor of $5$ \cite{najer2017}. In parallel, the suitability of the GeV as an efficient spin-photon interface should be confirmed through further characterization of the observed dark state and explicit measurement of the quantum efficiency in bulk diamond (which could be done using such a tuneable cavity). In the future, this platform could be easily adapted for coupling to novel emitters based on heavier group-IV elements such as the tin-vacancy (SnV) \cite{trusheim2018,rugar2019} and the lead-vacancy (PbV) centers \cite{trusheim2019, DitaliaTchernij2018}, which should exhibit longer spin coherence times. This work therefore encompasses several important steps toward realizing an efficient spin-photon interface for defect centers coupled to open microcavities. 

\begin{acknowledgments}
We would like to thank Marie-Jos\'ee Gour for assistance with membrane fabrication. We acknowledges support from  Canada Foundation for Innovation and Canada Research Chairs project 229003 and 231949,  Fonds de Recherche - Nature et Technologies FQRNT PR-253399, National Sciences and Engineering Research Council of Canada NSERC RGPIN 435554-13, l'Institut Transdisciplinaire d'Information Quantique (INTRIQ), the Danish Research Council through the Sapere Aude project DIMS, and the Danish National Research Foundation through the Center for Macroscopic Quantum States (bigQ, DNRF142). L. Childress is a CIFAR Fellow in the Quantum Information Science Program. Y. Fontana acknowledges support from SNSF Early Mobility Fellowship; CDRR was supported by a CONACYT fellowship; Y. Hi was supported by a MITACS Globalink internship.

\end{acknowledgments}
\nocite{hausmann2012, latawiec2015, appel2016, srim, chu2014, Brouri2000, chen2019, Neu2012, benedikter2017, janitz2015, Lukosz1977, Neyts1998, polerecky2000, reed1987, Albrecht2013, palyanov2015, rogers2014, katsidis2002,arnon1977, vandam2018, solin1970, iwasaki2015}

\bibliography{GeVbib}

\begin{thebibliography}{62}%
\makeatletter
\providecommand \@ifxundefined [1]{%
 \@ifx{#1\undefined}
}%
\providecommand \@ifnum [1]{%
 \ifnum #1\expandafter \@firstoftwo
 \else \expandafter \@secondoftwo
 \fi
}%
\providecommand \@ifx [1]{%
 \ifx #1\expandafter \@firstoftwo
 \else \expandafter \@secondoftwo
 \fi
}%
\providecommand \natexlab [1]{#1}%
\providecommand \enquote  [1]{``#1''}%
\providecommand \bibnamefont  [1]{#1}%
\providecommand \bibfnamefont [1]{#1}%
\providecommand \citenamefont [1]{#1}%
\providecommand \href@noop [0]{\@secondoftwo}%
\providecommand \href [0]{\begingroup \@sanitize@url \@href}%
\providecommand \@href[1]{\@@startlink{#1}\@@href}%
\providecommand \@@href[1]{\endgroup#1\@@endlink}%
\providecommand \@sanitize@url [0]{\catcode `\\12\catcode `\$12\catcode
  `\&12\catcode `\#12\catcode `\^12\catcode `\_12\catcode `\%12\relax}%
\providecommand \@@startlink[1]{}%
\providecommand \@@endlink[0]{}%
\providecommand \url  [0]{\begingroup\@sanitize@url \@url }%
\providecommand \@url [1]{\endgroup\@href {#1}{\urlprefix }}%
\providecommand \urlprefix  [0]{URL }%
\providecommand \Eprint [0]{\href }%
\providecommand \doibase [0]{https://doi.org/}%
\providecommand \selectlanguage [0]{\@gobble}%
\providecommand \bibinfo  [0]{\@secondoftwo}%
\providecommand \bibfield  [0]{\@secondoftwo}%
\providecommand \translation [1]{[#1]}%
\providecommand \BibitemOpen [0]{}%
\providecommand \bibitemStop [0]{}%
\providecommand \bibitemNoStop [0]{.\EOS\space}%
\providecommand \EOS [0]{\spacefactor3000\relax}%
\providecommand \BibitemShut  [1]{\csname bibitem#1\endcsname}%
\let\auto@bib@innerbib\@empty
\bibitem [{\citenamefont {Doherty}\ \emph {et~al.}(2013)\citenamefont
  {Doherty}, \citenamefont {Manson}, \citenamefont {Delaney}, \citenamefont
  {Jelezko}, \citenamefont {Wrachtrup},\ and\ \citenamefont
  {Hollenberg}}]{doherty2013}%
  \BibitemOpen
  \bibfield  {author} {\bibinfo {author} {\bibfnamefont {M.~W.}\ \bibnamefont
  {Doherty}}, \bibinfo {author} {\bibfnamefont {N.~B.}\ \bibnamefont {Manson}},
  \bibinfo {author} {\bibfnamefont {P.}~\bibnamefont {Delaney}}, \bibinfo
  {author} {\bibfnamefont {F.}~\bibnamefont {Jelezko}}, \bibinfo {author}
  {\bibfnamefont {J.}~\bibnamefont {Wrachtrup}},\ and\ \bibinfo {author}
  {\bibfnamefont {L.~C.}\ \bibnamefont {Hollenberg}},\ }\bibfield  {title}
  {\bibinfo {title} {{The nitrogen-vacancy colour centre in diamond}},\ }\href
  {https://doi.org/https://doi.org/10.1016/j.physrep.2013.02.001} {\bibfield
  {journal} {\bibinfo  {journal} {Physics Reports}\ }\textbf {\bibinfo {volume}
  {528}},\ \bibinfo {pages} {1 } (\bibinfo {year} {2013})}\BibitemShut
  {NoStop}%
\bibitem [{\citenamefont {Rose}\ \emph {et~al.}(2018)\citenamefont {Rose},
  \citenamefont {Huang}, \citenamefont {Zhang}, \citenamefont {Stevenson},
  \citenamefont {Tyryshkin}, \citenamefont {Sangtawesin}, \citenamefont
  {Srinivasan}, \citenamefont {Loudin}, \citenamefont {Markham}, \citenamefont
  {Edmonds} \emph {et~al.}}]{rose2018}%
  \BibitemOpen
  \bibfield  {author} {\bibinfo {author} {\bibfnamefont {B.~C.}\ \bibnamefont
  {Rose}}, \bibinfo {author} {\bibfnamefont {D.}~\bibnamefont {Huang}},
  \bibinfo {author} {\bibfnamefont {Z.-H.}\ \bibnamefont {Zhang}}, \bibinfo
  {author} {\bibfnamefont {P.}~\bibnamefont {Stevenson}}, \bibinfo {author}
  {\bibfnamefont {A.~M.}\ \bibnamefont {Tyryshkin}}, \bibinfo {author}
  {\bibfnamefont {S.}~\bibnamefont {Sangtawesin}}, \bibinfo {author}
  {\bibfnamefont {S.}~\bibnamefont {Srinivasan}}, \bibinfo {author}
  {\bibfnamefont {L.}~\bibnamefont {Loudin}}, \bibinfo {author} {\bibfnamefont
  {M.~L.}\ \bibnamefont {Markham}}, \bibinfo {author} {\bibfnamefont {A.~M.}\
  \bibnamefont {Edmonds}}, \emph {et~al.},\ }\bibfield  {title} {\bibinfo
  {title} {{Observation of an environmentally insensitive solid-state spin
  defect in diamond}},\ }\href {https://doi.org/10.1126/science.aao0290}
  {\bibfield  {journal} {\bibinfo  {journal} {Science}\ }\textbf {\bibinfo
  {volume} {361}},\ \bibinfo {pages} {60} (\bibinfo {year} {2018})}\BibitemShut
  {NoStop}%
\bibitem [{\citenamefont {Sukachev}\ \emph {et~al.}(2017)\citenamefont
  {Sukachev}, \citenamefont {Sipahigil}, \citenamefont {Nguyen}, \citenamefont
  {Bhaskar}, \citenamefont {Evans}, \citenamefont {Jelezko},\ and\
  \citenamefont {Lukin}}]{sukachev2017}%
  \BibitemOpen
  \bibfield  {author} {\bibinfo {author} {\bibfnamefont {D.~D.}\ \bibnamefont
  {Sukachev}}, \bibinfo {author} {\bibfnamefont {A.}~\bibnamefont {Sipahigil}},
  \bibinfo {author} {\bibfnamefont {C.~T.}\ \bibnamefont {Nguyen}}, \bibinfo
  {author} {\bibfnamefont {M.~K.}\ \bibnamefont {Bhaskar}}, \bibinfo {author}
  {\bibfnamefont {R.~E.}\ \bibnamefont {Evans}}, \bibinfo {author}
  {\bibfnamefont {F.}~\bibnamefont {Jelezko}},\ and\ \bibinfo {author}
  {\bibfnamefont {M.~D.}\ \bibnamefont {Lukin}},\ }\bibfield  {title} {\bibinfo
  {title} {{Silicon-Vacancy Spin Qubit in Diamond: A Quantum Memory Exceeding
  10 ms with Single-Shot State Readout}},\ }\href
  {https://doi.org/10.1103/PhysRevLett.119.223602} {\bibfield  {journal}
  {\bibinfo  {journal} {Physical Review Letters}\ }\textbf {\bibinfo {volume}
  {119}},\ \bibinfo {pages} {223602} (\bibinfo {year} {2017})}\BibitemShut
  {NoStop}%
\bibitem [{\citenamefont {Chang}\ \emph {et~al.}(2014)\citenamefont {Chang},
  \citenamefont {Vuleti{\'c}},\ and\ \citenamefont {Lukin}}]{chang2014}%
  \BibitemOpen
  \bibfield  {author} {\bibinfo {author} {\bibfnamefont {D.~E.}\ \bibnamefont
  {Chang}}, \bibinfo {author} {\bibfnamefont {V.}~\bibnamefont {Vuleti{\'c}}},\
  and\ \bibinfo {author} {\bibfnamefont {M.~D.}\ \bibnamefont {Lukin}},\
  }\bibfield  {title} {\bibinfo {title} {{Quantum nonlinear optics -€" photon
  by photon}},\ }\href {https://doi.org/10.1038/nphoton.2014.192} {\bibfield
  {journal} {\bibinfo  {journal} {Nature Photonics}\ }\textbf {\bibinfo
  {volume} {8}},\ \bibinfo {pages} {685} (\bibinfo {year} {2014})}\BibitemShut
  {NoStop}%
\bibitem [{\citenamefont {Kimble}(2008)}]{kimble2008}%
  \BibitemOpen
  \bibfield  {author} {\bibinfo {author} {\bibfnamefont {H.~J.}\ \bibnamefont
  {Kimble}},\ }\bibfield  {title} {\bibinfo {title} {The quantum internet},\
  }\href {https://doi.org/10.1038/nature07127} {\bibfield  {journal} {\bibinfo
  {journal} {Nature}\ }\textbf {\bibinfo {volume} {453}},\ \bibinfo {pages}
  {1023} (\bibinfo {year} {2008})}\BibitemShut {NoStop}%
\bibitem [{\citenamefont {Wehner}\ \emph {et~al.}(2018)\citenamefont {Wehner},
  \citenamefont {Elkouss},\ and\ \citenamefont {Hanson}}]{wehner2018}%
  \BibitemOpen
  \bibfield  {author} {\bibinfo {author} {\bibfnamefont {S.}~\bibnamefont
  {Wehner}}, \bibinfo {author} {\bibfnamefont {D.}~\bibnamefont {Elkouss}},\
  and\ \bibinfo {author} {\bibfnamefont {R.}~\bibnamefont {Hanson}},\
  }\bibfield  {title} {\bibinfo {title} {{Quantum internet: A vision for the
  road ahead}},\ }\href
  {https://science.sciencemag.org/content/362/6412/eaam9288} {\bibfield
  {journal} {\bibinfo  {journal} {Science}\ }\textbf {\bibinfo {volume}
  {362}},\ \bibinfo {pages} {eaam9288} (\bibinfo {year} {2018})}\BibitemShut
  {NoStop}%
\bibitem [{\citenamefont {Togan}\ \emph {et~al.}(2010)\citenamefont {Togan},
  \citenamefont {Chu}, \citenamefont {Trifonov}, \citenamefont {Jiang},
  \citenamefont {Maze}, \citenamefont {Childress}, \citenamefont {Dutt},
  \citenamefont {S{\o}rensen}, \citenamefont {Hemmer}, \citenamefont {Zibrov}
  \emph {et~al.}}]{togan2010}%
  \BibitemOpen
  \bibfield  {author} {\bibinfo {author} {\bibfnamefont {E.}~\bibnamefont
  {Togan}}, \bibinfo {author} {\bibfnamefont {Y.}~\bibnamefont {Chu}}, \bibinfo
  {author} {\bibfnamefont {A.}~\bibnamefont {Trifonov}}, \bibinfo {author}
  {\bibfnamefont {L.}~\bibnamefont {Jiang}}, \bibinfo {author} {\bibfnamefont
  {J.}~\bibnamefont {Maze}}, \bibinfo {author} {\bibfnamefont {L.}~\bibnamefont
  {Childress}}, \bibinfo {author} {\bibfnamefont {M.~G.}\ \bibnamefont {Dutt}},
  \bibinfo {author} {\bibfnamefont {A.~S.}\ \bibnamefont {S{\o}rensen}},
  \bibinfo {author} {\bibfnamefont {P.}~\bibnamefont {Hemmer}}, \bibinfo
  {author} {\bibfnamefont {A.~S.}\ \bibnamefont {Zibrov}}, \emph {et~al.},\
  }\bibfield  {title} {\bibinfo {title} {Quantum entanglement between an
  optical photon and a solid-state spin qubit},\ }\href
  {https://doi.org/10.1038/nature09256} {\bibfield  {journal} {\bibinfo
  {journal} {Nature}\ }\textbf {\bibinfo {volume} {466}},\ \bibinfo {pages}
  {730} (\bibinfo {year} {2010})}\BibitemShut {NoStop}%
\bibitem [{\citenamefont {Bernien}\ \emph {et~al.}(2012)\citenamefont
  {Bernien}, \citenamefont {Childress}, \citenamefont {Robledo}, \citenamefont
  {Markham}, \citenamefont {Twitchen},\ and\ \citenamefont
  {Hanson}}]{bernien2012}%
  \BibitemOpen
  \bibfield  {author} {\bibinfo {author} {\bibfnamefont {H.}~\bibnamefont
  {Bernien}}, \bibinfo {author} {\bibfnamefont {L.}~\bibnamefont {Childress}},
  \bibinfo {author} {\bibfnamefont {L.}~\bibnamefont {Robledo}}, \bibinfo
  {author} {\bibfnamefont {M.}~\bibnamefont {Markham}}, \bibinfo {author}
  {\bibfnamefont {D.}~\bibnamefont {Twitchen}},\ and\ \bibinfo {author}
  {\bibfnamefont {R.}~\bibnamefont {Hanson}},\ }\bibfield  {title} {\bibinfo
  {title} {{Two-Photon Quantum Interference from Separate Nitrogen Vacancy
  Centers in Diamond}},\ }\href
  {https://link.aps.org/doi/10.1103/PhysRevLett.108.043604} {\bibfield
  {journal} {\bibinfo  {journal} {Physical Review Letters}\ }\textbf {\bibinfo
  {volume} {108}},\ \bibinfo {pages} {043604} (\bibinfo {year}
  {2012})}\BibitemShut {NoStop}%
\bibitem [{\citenamefont {Bernien}\ \emph {et~al.}(2013)\citenamefont
  {Bernien}, \citenamefont {Hensen}, \citenamefont {Pfaff}, \citenamefont
  {Koolstra}, \citenamefont {Blok}, \citenamefont {Robledo}, \citenamefont
  {Taminiau}, \citenamefont {Markham}, \citenamefont {Twitchen}, \citenamefont
  {Childress} \emph {et~al.}}]{bernien2013}%
  \BibitemOpen
  \bibfield  {author} {\bibinfo {author} {\bibfnamefont {H.}~\bibnamefont
  {Bernien}}, \bibinfo {author} {\bibfnamefont {B.}~\bibnamefont {Hensen}},
  \bibinfo {author} {\bibfnamefont {W.}~\bibnamefont {Pfaff}}, \bibinfo
  {author} {\bibfnamefont {G.}~\bibnamefont {Koolstra}}, \bibinfo {author}
  {\bibfnamefont {M.}~\bibnamefont {Blok}}, \bibinfo {author} {\bibfnamefont
  {L.}~\bibnamefont {Robledo}}, \bibinfo {author} {\bibfnamefont
  {T.}~\bibnamefont {Taminiau}}, \bibinfo {author} {\bibfnamefont
  {M.}~\bibnamefont {Markham}}, \bibinfo {author} {\bibfnamefont
  {D.}~\bibnamefont {Twitchen}}, \bibinfo {author} {\bibfnamefont
  {L.}~\bibnamefont {Childress}}, \emph {et~al.},\ }\bibfield  {title}
  {\bibinfo {title} {Heralded entanglement between solid-state qubits separated
  by three metres},\ }\href {https://doi.org/10.1038/nature12016} {\bibfield
  {journal} {\bibinfo  {journal} {Nature}\ }\textbf {\bibinfo {volume} {497}},\
  \bibinfo {pages} {86} (\bibinfo {year} {2013})}\BibitemShut {NoStop}%
\bibitem [{\citenamefont {Hensen}\ \emph {et~al.}(2015)\citenamefont {Hensen},
  \citenamefont {Bernien}, \citenamefont {Dr{\'e}au}, \citenamefont {Reiserer},
  \citenamefont {Kalb}, \citenamefont {Blok}, \citenamefont {Ruitenberg},
  \citenamefont {Vermeulen}, \citenamefont {Schouten}, \citenamefont
  {Abell{\'a}n} \emph {et~al.}}]{hensen2015}%
  \BibitemOpen
  \bibfield  {author} {\bibinfo {author} {\bibfnamefont {B.}~\bibnamefont
  {Hensen}}, \bibinfo {author} {\bibfnamefont {H.}~\bibnamefont {Bernien}},
  \bibinfo {author} {\bibfnamefont {A.~E.}\ \bibnamefont {Dr{\'e}au}}, \bibinfo
  {author} {\bibfnamefont {A.}~\bibnamefont {Reiserer}}, \bibinfo {author}
  {\bibfnamefont {N.}~\bibnamefont {Kalb}}, \bibinfo {author} {\bibfnamefont
  {M.~S.}\ \bibnamefont {Blok}}, \bibinfo {author} {\bibfnamefont
  {J.}~\bibnamefont {Ruitenberg}}, \bibinfo {author} {\bibfnamefont {R.~F.}\
  \bibnamefont {Vermeulen}}, \bibinfo {author} {\bibfnamefont {R.~N.}\
  \bibnamefont {Schouten}}, \bibinfo {author} {\bibfnamefont {C.}~\bibnamefont
  {Abell{\'a}n}}, \emph {et~al.},\ }\bibfield  {title} {\bibinfo {title}
  {Loophole-free bell inequality violation using electron spins separated by
  1.3 kilometres},\ }\href {https://doi.org/10.1038/nature15759} {\bibfield
  {journal} {\bibinfo  {journal} {Nature}\ }\textbf {\bibinfo {volume} {526}},\
  \bibinfo {pages} {682} (\bibinfo {year} {2015})}\BibitemShut {NoStop}%
\bibitem [{\citenamefont {Pfaff}\ \emph {et~al.}(2014)\citenamefont {Pfaff},
  \citenamefont {Hensen}, \citenamefont {Bernien}, \citenamefont {van Dam},
  \citenamefont {Blok}, \citenamefont {Taminiau}, \citenamefont {Tiggelman},
  \citenamefont {Schouten}, \citenamefont {Markham}, \citenamefont {Twitchen}
  \emph {et~al.}}]{pfaff2014}%
  \BibitemOpen
  \bibfield  {author} {\bibinfo {author} {\bibfnamefont {W.}~\bibnamefont
  {Pfaff}}, \bibinfo {author} {\bibfnamefont {B.}~\bibnamefont {Hensen}},
  \bibinfo {author} {\bibfnamefont {H.}~\bibnamefont {Bernien}}, \bibinfo
  {author} {\bibfnamefont {S.~B.}\ \bibnamefont {van Dam}}, \bibinfo {author}
  {\bibfnamefont {M.~S.}\ \bibnamefont {Blok}}, \bibinfo {author}
  {\bibfnamefont {T.~H.}\ \bibnamefont {Taminiau}}, \bibinfo {author}
  {\bibfnamefont {M.~J.}\ \bibnamefont {Tiggelman}}, \bibinfo {author}
  {\bibfnamefont {R.~N.}\ \bibnamefont {Schouten}}, \bibinfo {author}
  {\bibfnamefont {M.}~\bibnamefont {Markham}}, \bibinfo {author} {\bibfnamefont
  {D.~J.}\ \bibnamefont {Twitchen}}, \emph {et~al.},\ }\bibfield  {title}
  {\bibinfo {title} {Unconditional quantum teleportation between distant
  solid-state quantum bits},\ }\href
  {https://science.sciencemag.org/content/345/6196/532} {\bibfield  {journal}
  {\bibinfo  {journal} {Science}\ }\textbf {\bibinfo {volume} {345}},\ \bibinfo
  {pages} {532} (\bibinfo {year} {2014})}\BibitemShut {NoStop}%
\bibitem [{\citenamefont {Maurer}\ \emph {et~al.}(2012)\citenamefont {Maurer},
  \citenamefont {Kucsko}, \citenamefont {Latta}, \citenamefont {Jiang},
  \citenamefont {Yao}, \citenamefont {Bennett}, \citenamefont {Pastawski},
  \citenamefont {Hunger}, \citenamefont {Chisholm}, \citenamefont {Markham}
  \emph {et~al.}}]{maurer2012}%
  \BibitemOpen
  \bibfield  {author} {\bibinfo {author} {\bibfnamefont {P.~C.}\ \bibnamefont
  {Maurer}}, \bibinfo {author} {\bibfnamefont {G.}~\bibnamefont {Kucsko}},
  \bibinfo {author} {\bibfnamefont {C.}~\bibnamefont {Latta}}, \bibinfo
  {author} {\bibfnamefont {L.}~\bibnamefont {Jiang}}, \bibinfo {author}
  {\bibfnamefont {N.~Y.}\ \bibnamefont {Yao}}, \bibinfo {author} {\bibfnamefont
  {S.~D.}\ \bibnamefont {Bennett}}, \bibinfo {author} {\bibfnamefont
  {F.}~\bibnamefont {Pastawski}}, \bibinfo {author} {\bibfnamefont
  {D.}~\bibnamefont {Hunger}}, \bibinfo {author} {\bibfnamefont
  {N.}~\bibnamefont {Chisholm}}, \bibinfo {author} {\bibfnamefont
  {M.}~\bibnamefont {Markham}}, \emph {et~al.},\ }\bibfield  {title} {\bibinfo
  {title} {Room-temperature quantum bit memory exceeding one second},\ }\href
  {https://science.sciencemag.org/content/336/6086/1283} {\bibfield  {journal}
  {\bibinfo  {journal} {Science}\ }\textbf {\bibinfo {volume} {336}},\ \bibinfo
  {pages} {1283} (\bibinfo {year} {2012})}\BibitemShut {NoStop}%
\bibitem [{\citenamefont {Kalb}\ \emph {et~al.}(2017)\citenamefont {Kalb},
  \citenamefont {Reiserer}, \citenamefont {Humphreys}, \citenamefont
  {Bakermans}, \citenamefont {Kamerling}, \citenamefont {Nickerson},
  \citenamefont {Benjamin}, \citenamefont {Twitchen}, \citenamefont {Markham},\
  and\ \citenamefont {Hanson}}]{kalb2017}%
  \BibitemOpen
  \bibfield  {author} {\bibinfo {author} {\bibfnamefont {N.}~\bibnamefont
  {Kalb}}, \bibinfo {author} {\bibfnamefont {A.~A.}\ \bibnamefont {Reiserer}},
  \bibinfo {author} {\bibfnamefont {P.~C.}\ \bibnamefont {Humphreys}}, \bibinfo
  {author} {\bibfnamefont {J.~J.}\ \bibnamefont {Bakermans}}, \bibinfo {author}
  {\bibfnamefont {S.~J.}\ \bibnamefont {Kamerling}}, \bibinfo {author}
  {\bibfnamefont {N.~H.}\ \bibnamefont {Nickerson}}, \bibinfo {author}
  {\bibfnamefont {S.~C.}\ \bibnamefont {Benjamin}}, \bibinfo {author}
  {\bibfnamefont {D.~J.}\ \bibnamefont {Twitchen}}, \bibinfo {author}
  {\bibfnamefont {M.}~\bibnamefont {Markham}},\ and\ \bibinfo {author}
  {\bibfnamefont {R.}~\bibnamefont {Hanson}},\ }\bibfield  {title} {\bibinfo
  {title} {Entanglement distillation between solid-state quantum network
  nodes},\ }\href {https://science.sciencemag.org/content/356/6341/928}
  {\bibfield  {journal} {\bibinfo  {journal} {Science}\ }\textbf {\bibinfo
  {volume} {356}},\ \bibinfo {pages} {928} (\bibinfo {year}
  {2017})}\BibitemShut {NoStop}%
\bibitem [{\citenamefont {Humphreys}\ \emph {et~al.}(2018)\citenamefont
  {Humphreys}, \citenamefont {Kalb}, \citenamefont {Morits}, \citenamefont
  {Schouten}, \citenamefont {Vermeulen}, \citenamefont {Twitchen},
  \citenamefont {Markham},\ and\ \citenamefont {Hanson}}]{humphreys2018}%
  \BibitemOpen
  \bibfield  {author} {\bibinfo {author} {\bibfnamefont {P.~C.}\ \bibnamefont
  {Humphreys}}, \bibinfo {author} {\bibfnamefont {N.}~\bibnamefont {Kalb}},
  \bibinfo {author} {\bibfnamefont {J.~P.}\ \bibnamefont {Morits}}, \bibinfo
  {author} {\bibfnamefont {R.~N.}\ \bibnamefont {Schouten}}, \bibinfo {author}
  {\bibfnamefont {R.~F.}\ \bibnamefont {Vermeulen}}, \bibinfo {author}
  {\bibfnamefont {D.~J.}\ \bibnamefont {Twitchen}}, \bibinfo {author}
  {\bibfnamefont {M.}~\bibnamefont {Markham}},\ and\ \bibinfo {author}
  {\bibfnamefont {R.}~\bibnamefont {Hanson}},\ }\bibfield  {title} {\bibinfo
  {title} {Deterministic delivery of remote entanglement on a quantum
  network},\ }\href {https://doi.org/10.1038/s41586-018-0200-5} {\bibfield
  {journal} {\bibinfo  {journal} {Nature}\ }\textbf {\bibinfo {volume} {558}},\
  \bibinfo {pages} {268} (\bibinfo {year} {2018})}\BibitemShut {NoStop}%
\bibitem [{\citenamefont {Faraon}\ \emph {et~al.}(2012)\citenamefont {Faraon},
  \citenamefont {Santori}, \citenamefont {Huang}, \citenamefont {Acosta},\ and\
  \citenamefont {Beausoleil}}]{faraon2012}%
  \BibitemOpen
  \bibfield  {author} {\bibinfo {author} {\bibfnamefont {A.}~\bibnamefont
  {Faraon}}, \bibinfo {author} {\bibfnamefont {C.}~\bibnamefont {Santori}},
  \bibinfo {author} {\bibfnamefont {Z.}~\bibnamefont {Huang}}, \bibinfo
  {author} {\bibfnamefont {V.~M.}\ \bibnamefont {Acosta}},\ and\ \bibinfo
  {author} {\bibfnamefont {R.~G.}\ \bibnamefont {Beausoleil}},\ }\bibfield
  {title} {\bibinfo {title} {{Coupling of Nitrogen-Vacancy Centers to Photonic
  Crystal Cavities in Monocrystalline Diamond}},\ }\href
  {https://link.aps.org/doi/10.1103/PhysRevLett.109.033604} {\bibfield
  {journal} {\bibinfo  {journal} {Physical Review Letters}\ }\textbf {\bibinfo
  {volume} {109}},\ \bibinfo {pages} {033604} (\bibinfo {year}
  {2012})}\BibitemShut {NoStop}%
\bibitem [{\citenamefont {Li}\ \emph {et~al.}(2015)\citenamefont {Li},
  \citenamefont {Schr{\"o}der}, \citenamefont {Chen}, \citenamefont {Walsh},
  \citenamefont {Bayn}, \citenamefont {Goldstein}, \citenamefont {Gaathon},
  \citenamefont {Trusheim}, \citenamefont {Lu}, \citenamefont {Mower} \emph
  {et~al.}}]{li2015}%
  \BibitemOpen
  \bibfield  {author} {\bibinfo {author} {\bibfnamefont {L.}~\bibnamefont
  {Li}}, \bibinfo {author} {\bibfnamefont {T.}~\bibnamefont {Schr{\"o}der}},
  \bibinfo {author} {\bibfnamefont {E.~H.}\ \bibnamefont {Chen}}, \bibinfo
  {author} {\bibfnamefont {M.}~\bibnamefont {Walsh}}, \bibinfo {author}
  {\bibfnamefont {I.}~\bibnamefont {Bayn}}, \bibinfo {author} {\bibfnamefont
  {J.}~\bibnamefont {Goldstein}}, \bibinfo {author} {\bibfnamefont
  {O.}~\bibnamefont {Gaathon}}, \bibinfo {author} {\bibfnamefont {M.~E.}\
  \bibnamefont {Trusheim}}, \bibinfo {author} {\bibfnamefont {M.}~\bibnamefont
  {Lu}}, \bibinfo {author} {\bibfnamefont {J.}~\bibnamefont {Mower}}, \emph
  {et~al.},\ }\bibfield  {title} {\bibinfo {title} {{Coherent spin control of a
  nanocavity-enhanced qubit in diamond}},\ }\href
  {https://doi.org/10.1038/ncomms7173} {\bibfield  {journal} {\bibinfo
  {journal} {Nature Communications}\ }\textbf {\bibinfo {volume} {6}},\
  \bibinfo {pages} {6173} (\bibinfo {year} {2015})}\BibitemShut {NoStop}%
\bibitem [{\citenamefont {Riedel}\ \emph {et~al.}(2017)\citenamefont {Riedel},
  \citenamefont {S{\"o}llner}, \citenamefont {Shields}, \citenamefont
  {Starosielec}, \citenamefont {Appel}, \citenamefont {Neu}, \citenamefont
  {Maletinsky},\ and\ \citenamefont {Warburton}}]{riedel2017}%
  \BibitemOpen
  \bibfield  {author} {\bibinfo {author} {\bibfnamefont {D.}~\bibnamefont
  {Riedel}}, \bibinfo {author} {\bibfnamefont {I.}~\bibnamefont {S{\"o}llner}},
  \bibinfo {author} {\bibfnamefont {B.~J.}\ \bibnamefont {Shields}}, \bibinfo
  {author} {\bibfnamefont {S.}~\bibnamefont {Starosielec}}, \bibinfo {author}
  {\bibfnamefont {P.}~\bibnamefont {Appel}}, \bibinfo {author} {\bibfnamefont
  {E.}~\bibnamefont {Neu}}, \bibinfo {author} {\bibfnamefont {P.}~\bibnamefont
  {Maletinsky}},\ and\ \bibinfo {author} {\bibfnamefont {R.~J.}\ \bibnamefont
  {Warburton}},\ }\bibfield  {title} {\bibinfo {title} {{Deterministic
  Enhancement of Coherent Photon Generation from a Nitrogen-Vacancy Center in
  Ultrapure Diamond}},\ }\href
  {https://link.aps.org/doi/10.1103/PhysRevX.7.031040} {\bibfield  {journal}
  {\bibinfo  {journal} {Physical Review X}\ }\textbf {\bibinfo {volume} {7}},\
  \bibinfo {pages} {031040} (\bibinfo {year} {2017})}\BibitemShut {NoStop}%
\bibitem [{\citenamefont {Hepp}\ \emph {et~al.}(2014)\citenamefont {Hepp},
  \citenamefont {M{\"u}ller}, \citenamefont {Waselowski}, \citenamefont
  {Becker}, \citenamefont {Pingault}, \citenamefont {Sternschulte},
  \citenamefont {Steinm{\"u}ller-Nethl}, \citenamefont {Gali}, \citenamefont
  {Maze}, \citenamefont {Atat{\"u}re} \emph {et~al.}}]{hepp2014}%
  \BibitemOpen
  \bibfield  {author} {\bibinfo {author} {\bibfnamefont {C.}~\bibnamefont
  {Hepp}}, \bibinfo {author} {\bibfnamefont {T.}~\bibnamefont {M{\"u}ller}},
  \bibinfo {author} {\bibfnamefont {V.}~\bibnamefont {Waselowski}}, \bibinfo
  {author} {\bibfnamefont {J.~N.}\ \bibnamefont {Becker}}, \bibinfo {author}
  {\bibfnamefont {B.}~\bibnamefont {Pingault}}, \bibinfo {author}
  {\bibfnamefont {H.}~\bibnamefont {Sternschulte}}, \bibinfo {author}
  {\bibfnamefont {D.}~\bibnamefont {Steinm{\"u}ller-Nethl}}, \bibinfo {author}
  {\bibfnamefont {A.}~\bibnamefont {Gali}}, \bibinfo {author} {\bibfnamefont
  {J.~R.}\ \bibnamefont {Maze}}, \bibinfo {author} {\bibfnamefont
  {M.}~\bibnamefont {Atat{\"u}re}}, \emph {et~al.},\ }\bibfield  {title}
  {\bibinfo {title} {{Electronic Structure of the Silicon Vacancy Color Center
  in Diamond}},\ }\href
  {https://link.aps.org/doi/10.1103/PhysRevLett.112.036405} {\bibfield
  {journal} {\bibinfo  {journal} {Physical Review Letters}\ }\textbf {\bibinfo
  {volume} {112}},\ \bibinfo {pages} {036405} (\bibinfo {year}
  {2014})}\BibitemShut {NoStop}%
\bibitem [{\citenamefont {Rogers}\ \emph {et~al.}(2014)\citenamefont {Rogers},
  \citenamefont {Jahnke}, \citenamefont {Doherty}, \citenamefont {Dietrich},
  \citenamefont {McGuinness}, \citenamefont {M{\"u}ller}, \citenamefont
  {Teraji}, \citenamefont {Sumiya}, \citenamefont {Isoya}, \citenamefont
  {Manson} \emph {et~al.}}]{rogers2014}%
  \BibitemOpen
  \bibfield  {author} {\bibinfo {author} {\bibfnamefont {L.~J.}\ \bibnamefont
  {Rogers}}, \bibinfo {author} {\bibfnamefont {K.~D.}\ \bibnamefont {Jahnke}},
  \bibinfo {author} {\bibfnamefont {M.~W.}\ \bibnamefont {Doherty}}, \bibinfo
  {author} {\bibfnamefont {A.}~\bibnamefont {Dietrich}}, \bibinfo {author}
  {\bibfnamefont {L.~P.}\ \bibnamefont {McGuinness}}, \bibinfo {author}
  {\bibfnamefont {C.}~\bibnamefont {M{\"u}ller}}, \bibinfo {author}
  {\bibfnamefont {T.}~\bibnamefont {Teraji}}, \bibinfo {author} {\bibfnamefont
  {H.}~\bibnamefont {Sumiya}}, \bibinfo {author} {\bibfnamefont
  {J.}~\bibnamefont {Isoya}}, \bibinfo {author} {\bibfnamefont {N.~B.}\
  \bibnamefont {Manson}}, \emph {et~al.},\ }\bibfield  {title} {\bibinfo
  {title} {Electronic structure of the negatively charged silicon-vacancy
  center in diamond},\ }\href
  {https://link.aps.org/doi/10.1103/PhysRevB.89.235101} {\bibfield  {journal}
  {\bibinfo  {journal} {Physical Review B}\ }\textbf {\bibinfo {volume} {89}},\
  \bibinfo {pages} {235101} (\bibinfo {year} {2014})}\BibitemShut {NoStop}%
\bibitem [{\citenamefont {Siyushev}\ \emph {et~al.}(2017)\citenamefont
  {Siyushev}, \citenamefont {Metsch}, \citenamefont {Ijaz}, \citenamefont
  {Binder}, \citenamefont {Bhaskar}, \citenamefont {Sukachev}, \citenamefont
  {Sipahigil}, \citenamefont {Evans}, \citenamefont {Nguyen}, \citenamefont
  {Lukin} \emph {et~al.}}]{siyushev2017}%
  \BibitemOpen
  \bibfield  {author} {\bibinfo {author} {\bibfnamefont {P.}~\bibnamefont
  {Siyushev}}, \bibinfo {author} {\bibfnamefont {M.~H.}\ \bibnamefont
  {Metsch}}, \bibinfo {author} {\bibfnamefont {A.}~\bibnamefont {Ijaz}},
  \bibinfo {author} {\bibfnamefont {J.~M.}\ \bibnamefont {Binder}}, \bibinfo
  {author} {\bibfnamefont {M.~K.}\ \bibnamefont {Bhaskar}}, \bibinfo {author}
  {\bibfnamefont {D.~D.}\ \bibnamefont {Sukachev}}, \bibinfo {author}
  {\bibfnamefont {A.}~\bibnamefont {Sipahigil}}, \bibinfo {author}
  {\bibfnamefont {R.~E.}\ \bibnamefont {Evans}}, \bibinfo {author}
  {\bibfnamefont {C.~T.}\ \bibnamefont {Nguyen}}, \bibinfo {author}
  {\bibfnamefont {M.~D.}\ \bibnamefont {Lukin}}, \emph {et~al.},\ }\bibfield
  {title} {\bibinfo {title} {Optical and microwave control of germanium-vacancy
  center spins in diamond},\ }\href
  {https://link.aps.org/doi/10.1103/PhysRevB.96.081201} {\bibfield  {journal}
  {\bibinfo  {journal} {Physical Review B}\ }\textbf {\bibinfo {volume} {96}},\
  \bibinfo {pages} {081201} (\bibinfo {year} {2017})}\BibitemShut {NoStop}%
\bibitem [{\citenamefont {Iwasaki}\ \emph {et~al.}(2015)\citenamefont
  {Iwasaki}, \citenamefont {Ishibashi}, \citenamefont {Miyamoto}, \citenamefont
  {Doi}, \citenamefont {Kobayashi}, \citenamefont {Miyazaki}, \citenamefont
  {Tahara}, \citenamefont {Jahnke}, \citenamefont {Rogers}, \citenamefont
  {Naydenov} \emph {et~al.}}]{iwasaki2015}%
  \BibitemOpen
  \bibfield  {author} {\bibinfo {author} {\bibfnamefont {T.}~\bibnamefont
  {Iwasaki}}, \bibinfo {author} {\bibfnamefont {F.}~\bibnamefont {Ishibashi}},
  \bibinfo {author} {\bibfnamefont {Y.}~\bibnamefont {Miyamoto}}, \bibinfo
  {author} {\bibfnamefont {Y.}~\bibnamefont {Doi}}, \bibinfo {author}
  {\bibfnamefont {S.}~\bibnamefont {Kobayashi}}, \bibinfo {author}
  {\bibfnamefont {T.}~\bibnamefont {Miyazaki}}, \bibinfo {author}
  {\bibfnamefont {K.}~\bibnamefont {Tahara}}, \bibinfo {author} {\bibfnamefont
  {K.~D.}\ \bibnamefont {Jahnke}}, \bibinfo {author} {\bibfnamefont {L.~J.}\
  \bibnamefont {Rogers}}, \bibinfo {author} {\bibfnamefont {B.}~\bibnamefont
  {Naydenov}}, \emph {et~al.},\ }\bibfield  {title} {\bibinfo {title}
  {{Germanium-Vacancy Single Color Centers in Diamond}},\ }\href
  {https://doi.org/10.1038/srep12882} {\bibfield  {journal} {\bibinfo
  {journal} {Scientific Reports}\ }\textbf {\bibinfo {volume} {5}},\ \bibinfo
  {pages} {12882} (\bibinfo {year} {2015})}\BibitemShut {NoStop}%
\bibitem [{\citenamefont {Sohn}\ \emph {et~al.}(2018)\citenamefont {Sohn},
  \citenamefont {Meesala}, \citenamefont {Pingault}, \citenamefont {Atikian},
  \citenamefont {Holzgrafe}, \citenamefont {G{\"u}ndo{\u{g}}an}, \citenamefont
  {Stavrakas}, \citenamefont {Stanley}, \citenamefont {Sipahigil},
  \citenamefont {Choi} \emph {et~al.}}]{sohn2018}%
  \BibitemOpen
  \bibfield  {author} {\bibinfo {author} {\bibfnamefont {Y.-I.}\ \bibnamefont
  {Sohn}}, \bibinfo {author} {\bibfnamefont {S.}~\bibnamefont {Meesala}},
  \bibinfo {author} {\bibfnamefont {B.}~\bibnamefont {Pingault}}, \bibinfo
  {author} {\bibfnamefont {H.~A.}\ \bibnamefont {Atikian}}, \bibinfo {author}
  {\bibfnamefont {J.}~\bibnamefont {Holzgrafe}}, \bibinfo {author}
  {\bibfnamefont {M.}~\bibnamefont {G{\"u}ndo{\u{g}}an}}, \bibinfo {author}
  {\bibfnamefont {C.}~\bibnamefont {Stavrakas}}, \bibinfo {author}
  {\bibfnamefont {M.~J.}\ \bibnamefont {Stanley}}, \bibinfo {author}
  {\bibfnamefont {A.}~\bibnamefont {Sipahigil}}, \bibinfo {author}
  {\bibfnamefont {J.}~\bibnamefont {Choi}}, \emph {et~al.},\ }\bibfield
  {title} {\bibinfo {title} {Controlling the coherence of a diamond spin qubit
  through its strain environment},\ }\href
  {https://doi.org/10.1038/s41467-018-04340-3} {\bibfield  {journal} {\bibinfo
  {journal} {Nature Communications}\ }\textbf {\bibinfo {volume} {9}},\
  \bibinfo {pages} {2012} (\bibinfo {year} {2018})}\BibitemShut {NoStop}%
\bibitem [{\citenamefont {Thiering}\ and\ \citenamefont
  {Gali}(2018)}]{Thiering2017}%
  \BibitemOpen
  \bibfield  {author} {\bibinfo {author} {\bibfnamefont {G.~m.~H.}\
  \bibnamefont {Thiering}}\ and\ \bibinfo {author} {\bibfnamefont
  {A.}~\bibnamefont {Gali}},\ }\bibfield  {title} {\bibinfo {title} {{Ab Initio
  Magneto-Optical Spectrum of Group-IV Vacancy Color Centers in Diamond}},\
  }\href {https://doi.org/10.1103/PhysRevX.8.021063} {\bibfield  {journal}
  {\bibinfo  {journal} {Physical Review X}\ }\textbf {\bibinfo {volume} {8}},\
  \bibinfo {pages} {021063} (\bibinfo {year} {2018})}\BibitemShut {NoStop}%
\bibitem [{\citenamefont {Evans}\ \emph {et~al.}(2016)\citenamefont {Evans},
  \citenamefont {Sipahigil}, \citenamefont {Sukachev}, \citenamefont {Zibrov},\
  and\ \citenamefont {Lukin}}]{evans2016}%
  \BibitemOpen
  \bibfield  {author} {\bibinfo {author} {\bibfnamefont {R.~E.}\ \bibnamefont
  {Evans}}, \bibinfo {author} {\bibfnamefont {A.}~\bibnamefont {Sipahigil}},
  \bibinfo {author} {\bibfnamefont {D.~D.}\ \bibnamefont {Sukachev}}, \bibinfo
  {author} {\bibfnamefont {A.~S.}\ \bibnamefont {Zibrov}},\ and\ \bibinfo
  {author} {\bibfnamefont {M.~D.}\ \bibnamefont {Lukin}},\ }\bibfield  {title}
  {\bibinfo {title} {{Narrow-Linewidth Homogeneous Optical Emitters in Diamond
  Nanostructures via Silicon Ion Implantation}},\ }\href
  {https://link.aps.org/doi/10.1103/PhysRevApplied.5.044010} {\bibfield
  {journal} {\bibinfo  {journal} {Physical Review Applied}\ }\textbf {\bibinfo
  {volume} {5}},\ \bibinfo {pages} {044010} (\bibinfo {year}
  {2016})}\BibitemShut {NoStop}%
\bibitem [{\citenamefont {Bhaskar}\ \emph {et~al.}(2017)\citenamefont
  {Bhaskar}, \citenamefont {Sukachev}, \citenamefont {Sipahigil}, \citenamefont
  {Evans}, \citenamefont {Burek}, \citenamefont {Nguyen}, \citenamefont
  {Rogers}, \citenamefont {Siyushev}, \citenamefont {Metsch}, \citenamefont
  {Park} \emph {et~al.}}]{Bhaskar2017}%
  \BibitemOpen
  \bibfield  {author} {\bibinfo {author} {\bibfnamefont {M.~K.}\ \bibnamefont
  {Bhaskar}}, \bibinfo {author} {\bibfnamefont {D.~D.}\ \bibnamefont
  {Sukachev}}, \bibinfo {author} {\bibfnamefont {A.}~\bibnamefont {Sipahigil}},
  \bibinfo {author} {\bibfnamefont {R.~E.}\ \bibnamefont {Evans}}, \bibinfo
  {author} {\bibfnamefont {M.~J.}\ \bibnamefont {Burek}}, \bibinfo {author}
  {\bibfnamefont {C.~T.}\ \bibnamefont {Nguyen}}, \bibinfo {author}
  {\bibfnamefont {L.~J.}\ \bibnamefont {Rogers}}, \bibinfo {author}
  {\bibfnamefont {P.}~\bibnamefont {Siyushev}}, \bibinfo {author}
  {\bibfnamefont {M.~H.}\ \bibnamefont {Metsch}}, \bibinfo {author}
  {\bibfnamefont {H.}~\bibnamefont {Park}}, \emph {et~al.},\ }\bibfield
  {title} {\bibinfo {title} {{Quantum Nonlinear Optics with a Germanium-Vacancy
  Color Center in a Nanoscale Diamond Waveguide}},\ }\href
  {https://doi.org/10.1103/PhysRevLett.118.223603} {\bibfield  {journal}
  {\bibinfo  {journal} {Physical Review Letters}\ }\textbf {\bibinfo {volume}
  {118}},\ \bibinfo {pages} {223603} (\bibinfo {year} {2017})}\BibitemShut
  {NoStop}%
\bibitem [{\citenamefont {Wan}\ \emph {et~al.}(2019)\citenamefont {Wan},
  \citenamefont {Lu}, \citenamefont {Chen}, \citenamefont {Walsh},
  \citenamefont {Trusheim}, \citenamefont {De~Santis}, \citenamefont {Bersin},
  \citenamefont {Harris}, \citenamefont {Mouradian}, \citenamefont {Christen}
  \emph {et~al.}}]{wan2019}%
  \BibitemOpen
  \bibfield  {author} {\bibinfo {author} {\bibfnamefont {N.~H.}\ \bibnamefont
  {Wan}}, \bibinfo {author} {\bibfnamefont {T.-J.}\ \bibnamefont {Lu}},
  \bibinfo {author} {\bibfnamefont {K.~C.}\ \bibnamefont {Chen}}, \bibinfo
  {author} {\bibfnamefont {M.~P.}\ \bibnamefont {Walsh}}, \bibinfo {author}
  {\bibfnamefont {M.~E.}\ \bibnamefont {Trusheim}}, \bibinfo {author}
  {\bibfnamefont {L.}~\bibnamefont {De~Santis}}, \bibinfo {author}
  {\bibfnamefont {E.~A.}\ \bibnamefont {Bersin}}, \bibinfo {author}
  {\bibfnamefont {I.~B.}\ \bibnamefont {Harris}}, \bibinfo {author}
  {\bibfnamefont {S.~L.}\ \bibnamefont {Mouradian}}, \bibinfo {author}
  {\bibfnamefont {I.~R.}\ \bibnamefont {Christen}}, \emph {et~al.},\ }\bibfield
   {title} {\bibinfo {title} {Large-scale integration of near-indistinguishable
  artificial atoms in hybrid photonic circuits},\ }\href@noop {} {\bibfield
  {journal} {\bibinfo  {journal} {arXiv preprint arXiv:1911.05265}\ } (\bibinfo
  {year} {2019})}\BibitemShut {NoStop}%
\bibitem [{\citenamefont {Evans}\ \emph {et~al.}(2018)\citenamefont {Evans},
  \citenamefont {Bhaskar}, \citenamefont {Sukachev}, \citenamefont {Nguyen},
  \citenamefont {Sipahigil}, \citenamefont {Burek}, \citenamefont {Machielse},
  \citenamefont {Zhang}, \citenamefont {Zibrov}, \citenamefont {Bielejec} \emph
  {et~al.}}]{evans2018}%
  \BibitemOpen
  \bibfield  {author} {\bibinfo {author} {\bibfnamefont {R.~E.}\ \bibnamefont
  {Evans}}, \bibinfo {author} {\bibfnamefont {M.~K.}\ \bibnamefont {Bhaskar}},
  \bibinfo {author} {\bibfnamefont {D.~D.}\ \bibnamefont {Sukachev}}, \bibinfo
  {author} {\bibfnamefont {C.~T.}\ \bibnamefont {Nguyen}}, \bibinfo {author}
  {\bibfnamefont {A.}~\bibnamefont {Sipahigil}}, \bibinfo {author}
  {\bibfnamefont {M.~J.}\ \bibnamefont {Burek}}, \bibinfo {author}
  {\bibfnamefont {B.}~\bibnamefont {Machielse}}, \bibinfo {author}
  {\bibfnamefont {G.~H.}\ \bibnamefont {Zhang}}, \bibinfo {author}
  {\bibfnamefont {A.~S.}\ \bibnamefont {Zibrov}}, \bibinfo {author}
  {\bibfnamefont {E.}~\bibnamefont {Bielejec}}, \emph {et~al.},\ }\bibfield
  {title} {\bibinfo {title} {Photon-mediated interactions between quantum
  emitters in a diamond nanocavity},\ }\href
  {https://science.sciencemag.org/content/362/6415/662} {\bibfield  {journal}
  {\bibinfo  {journal} {Science}\ }\textbf {\bibinfo {volume} {362}},\ \bibinfo
  {pages} {662} (\bibinfo {year} {2018})}\BibitemShut {NoStop}%
\bibitem [{\citenamefont {Sipahigil}\ \emph {et~al.}(2016)\citenamefont
  {Sipahigil}, \citenamefont {Evans}, \citenamefont {Sukachev}, \citenamefont
  {Burek}, \citenamefont {Borregaard}, \citenamefont {Bhaskar}, \citenamefont
  {Nguyen}, \citenamefont {Pacheco}, \citenamefont {Atikian}, \citenamefont
  {Meuwly} \emph {et~al.}}]{sipahigil2016}%
  \BibitemOpen
  \bibfield  {author} {\bibinfo {author} {\bibfnamefont {A.}~\bibnamefont
  {Sipahigil}}, \bibinfo {author} {\bibfnamefont {R.~E.}\ \bibnamefont
  {Evans}}, \bibinfo {author} {\bibfnamefont {D.~D.}\ \bibnamefont {Sukachev}},
  \bibinfo {author} {\bibfnamefont {M.~J.}\ \bibnamefont {Burek}}, \bibinfo
  {author} {\bibfnamefont {J.}~\bibnamefont {Borregaard}}, \bibinfo {author}
  {\bibfnamefont {M.~K.}\ \bibnamefont {Bhaskar}}, \bibinfo {author}
  {\bibfnamefont {C.~T.}\ \bibnamefont {Nguyen}}, \bibinfo {author}
  {\bibfnamefont {J.~L.}\ \bibnamefont {Pacheco}}, \bibinfo {author}
  {\bibfnamefont {H.~A.}\ \bibnamefont {Atikian}}, \bibinfo {author}
  {\bibfnamefont {C.}~\bibnamefont {Meuwly}}, \emph {et~al.},\ }\bibfield
  {title} {\bibinfo {title} {An integrated diamond nanophotonics platform for
  quantum-optical networks},\ }\href
  {https://science.sciencemag.org/content/354/6314/847} {\bibfield  {journal}
  {\bibinfo  {journal} {Science}\ }\textbf {\bibinfo {volume} {354}},\ \bibinfo
  {pages} {847} (\bibinfo {year} {2016})}\BibitemShut {NoStop}%
\bibitem [{\citenamefont {Nguyen}\ \emph
  {et~al.}(2019{\natexlab{a}})\citenamefont {Nguyen}, \citenamefont {Sukachev},
  \citenamefont {Bhaskar}, \citenamefont {Machielse}, \citenamefont {Levonian},
  \citenamefont {Knall}, \citenamefont {Stroganov}, \citenamefont {Chia},
  \citenamefont {Burek}, \citenamefont {Riedinger} \emph
  {et~al.}}]{nguyen2019}%
  \BibitemOpen
  \bibfield  {author} {\bibinfo {author} {\bibfnamefont {C.}~\bibnamefont
  {Nguyen}}, \bibinfo {author} {\bibfnamefont {D.}~\bibnamefont {Sukachev}},
  \bibinfo {author} {\bibfnamefont {M.}~\bibnamefont {Bhaskar}}, \bibinfo
  {author} {\bibfnamefont {B.}~\bibnamefont {Machielse}}, \bibinfo {author}
  {\bibfnamefont {D.}~\bibnamefont {Levonian}}, \bibinfo {author}
  {\bibfnamefont {E.}~\bibnamefont {Knall}}, \bibinfo {author} {\bibfnamefont
  {P.}~\bibnamefont {Stroganov}}, \bibinfo {author} {\bibfnamefont
  {C.}~\bibnamefont {Chia}}, \bibinfo {author} {\bibfnamefont {M.}~\bibnamefont
  {Burek}}, \bibinfo {author} {\bibfnamefont {R.}~\bibnamefont {Riedinger}},
  \emph {et~al.},\ }\bibfield  {title} {\bibinfo {title} {An integrated
  nanophotonic quantum register based on silicon-vacancy spins in diamond},\
  }\href {https://link.aps.org/doi/10.1103/PhysRevB.100.165428} {\bibfield
  {journal} {\bibinfo  {journal} {Physical Review B}\ }\textbf {\bibinfo
  {volume} {100}},\ \bibinfo {pages} {165428} (\bibinfo {year}
  {2019}{\natexlab{a}})}\BibitemShut {NoStop}%
\bibitem [{\citenamefont {Ruf}\ \emph {et~al.}(2019)\citenamefont {Ruf},
  \citenamefont {IJspeert}, \citenamefont {van Dam}, \citenamefont {de~Jong},
  \citenamefont {van~den Berg}, \citenamefont {Evers},\ and\ \citenamefont
  {Hanson}}]{ruf2019}%
  \BibitemOpen
  \bibfield  {author} {\bibinfo {author} {\bibfnamefont {M.}~\bibnamefont
  {Ruf}}, \bibinfo {author} {\bibfnamefont {M.}~\bibnamefont {IJspeert}},
  \bibinfo {author} {\bibfnamefont {S.}~\bibnamefont {van Dam}}, \bibinfo
  {author} {\bibfnamefont {N.}~\bibnamefont {de~Jong}}, \bibinfo {author}
  {\bibfnamefont {H.}~\bibnamefont {van~den Berg}}, \bibinfo {author}
  {\bibfnamefont {G.}~\bibnamefont {Evers}},\ and\ \bibinfo {author}
  {\bibfnamefont {R.}~\bibnamefont {Hanson}},\ }\bibfield  {title} {\bibinfo
  {title} {{Optically Coherent Nitrogen-Vacancy Centers in Micrometer-Thin
  Etched Diamond Membranes}},\ }\href
  {https://pubs.acs.org/doi/10.1021/acs.nanolett.9b01316} {\bibfield  {journal}
  {\bibinfo  {journal} {Nano Letters}\ }\textbf {\bibinfo {volume} {19}},\
  \bibinfo {pages} {3987} (\bibinfo {year} {2019})}\BibitemShut {NoStop}%
\bibitem [{\citenamefont {Janitz}\ \emph {et~al.}(2015)\citenamefont {Janitz},
  \citenamefont {Ruf}, \citenamefont {Dimock}, \citenamefont {Bourassa},
  \citenamefont {Sankey},\ and\ \citenamefont {Childress}}]{janitz2015}%
  \BibitemOpen
  \bibfield  {author} {\bibinfo {author} {\bibfnamefont {E.}~\bibnamefont
  {Janitz}}, \bibinfo {author} {\bibfnamefont {M.}~\bibnamefont {Ruf}},
  \bibinfo {author} {\bibfnamefont {M.}~\bibnamefont {Dimock}}, \bibinfo
  {author} {\bibfnamefont {A.}~\bibnamefont {Bourassa}}, \bibinfo {author}
  {\bibfnamefont {J.}~\bibnamefont {Sankey}},\ and\ \bibinfo {author}
  {\bibfnamefont {L.}~\bibnamefont {Childress}},\ }\bibfield  {title} {\bibinfo
  {title} {{Fabry-Perot microcavity for diamond-based photonics}},\ }\href
  {https://link.aps.org/doi/10.1103/PhysRevA.92.043844} {\bibfield  {journal}
  {\bibinfo  {journal} {Physical Review A}\ }\textbf {\bibinfo {volume} {92}},\
  \bibinfo {pages} {043844} (\bibinfo {year} {2015})}\BibitemShut {NoStop}%
\bibitem [{\citenamefont {H{\"a}u{\ss}ler}\ \emph {et~al.}(2019)\citenamefont
  {H{\"a}u{\ss}ler}, \citenamefont {Benedikter}, \citenamefont {Bray},
  \citenamefont {Regan}, \citenamefont {Dietrich}, \citenamefont {Twamley},
  \citenamefont {Aharonovich}, \citenamefont {Hunger},\ and\ \citenamefont
  {Kubanek}}]{haussler2019}%
  \BibitemOpen
  \bibfield  {author} {\bibinfo {author} {\bibfnamefont {S.}~\bibnamefont
  {H{\"a}u{\ss}ler}}, \bibinfo {author} {\bibfnamefont {J.}~\bibnamefont
  {Benedikter}}, \bibinfo {author} {\bibfnamefont {K.}~\bibnamefont {Bray}},
  \bibinfo {author} {\bibfnamefont {B.}~\bibnamefont {Regan}}, \bibinfo
  {author} {\bibfnamefont {A.}~\bibnamefont {Dietrich}}, \bibinfo {author}
  {\bibfnamefont {J.}~\bibnamefont {Twamley}}, \bibinfo {author} {\bibfnamefont
  {I.}~\bibnamefont {Aharonovich}}, \bibinfo {author} {\bibfnamefont
  {D.}~\bibnamefont {Hunger}},\ and\ \bibinfo {author} {\bibfnamefont
  {A.}~\bibnamefont {Kubanek}},\ }\bibfield  {title} {\bibinfo {title}
  {{Diamond photonics platform based on silicon vacancy centers in a
  single-crystal diamond membrane and a fiber cavity}},\ }\href
  {https://link.aps.org/doi/10.1103/PhysRevB.99.165310} {\bibfield  {journal}
  {\bibinfo  {journal} {Physical Review B}\ }\textbf {\bibinfo {volume} {99}},\
  \bibinfo {pages} {165310} (\bibinfo {year} {2019})}\BibitemShut {NoStop}%
\bibitem [{\citenamefont {Albrecht}\ \emph {et~al.}(2013)\citenamefont
  {Albrecht}, \citenamefont {Bommer}, \citenamefont {Deutsch}, \citenamefont
  {Reichel},\ and\ \citenamefont {Becher}}]{Albrecht2013}%
  \BibitemOpen
  \bibfield  {author} {\bibinfo {author} {\bibfnamefont {R.}~\bibnamefont
  {Albrecht}}, \bibinfo {author} {\bibfnamefont {A.}~\bibnamefont {Bommer}},
  \bibinfo {author} {\bibfnamefont {C.}~\bibnamefont {Deutsch}}, \bibinfo
  {author} {\bibfnamefont {J.}~\bibnamefont {Reichel}},\ and\ \bibinfo {author}
  {\bibfnamefont {C.}~\bibnamefont {Becher}},\ }\bibfield  {title} {\bibinfo
  {title} {{Coupling of a Single Nitrogen-Vacancy Center in Diamond to a
  Fiber-Based Microcavity}},\ }\href
  {https://link.aps.org/doi/10.1103/PhysRevLett.110.243602} {\bibfield
  {journal} {\bibinfo  {journal} {Physical Review Letters}\ }\textbf {\bibinfo
  {volume} {110}} (\bibinfo {year} {2013})}\BibitemShut {NoStop}%
\bibitem [{\citenamefont {Benedikter}\ \emph {et~al.}(2017)\citenamefont
  {Benedikter}, \citenamefont {Kaupp}, \citenamefont {H{\"u}mmer},
  \citenamefont {Liang}, \citenamefont {Bommer}, \citenamefont {Becher},
  \citenamefont {Krueger}, \citenamefont {Smith}, \citenamefont {H{\"a}nsch},\
  and\ \citenamefont {Hunger}}]{benedikter2017}%
  \BibitemOpen
  \bibfield  {author} {\bibinfo {author} {\bibfnamefont {J.}~\bibnamefont
  {Benedikter}}, \bibinfo {author} {\bibfnamefont {H.}~\bibnamefont {Kaupp}},
  \bibinfo {author} {\bibfnamefont {T.}~\bibnamefont {H{\"u}mmer}}, \bibinfo
  {author} {\bibfnamefont {Y.}~\bibnamefont {Liang}}, \bibinfo {author}
  {\bibfnamefont {A.}~\bibnamefont {Bommer}}, \bibinfo {author} {\bibfnamefont
  {C.}~\bibnamefont {Becher}}, \bibinfo {author} {\bibfnamefont
  {A.}~\bibnamefont {Krueger}}, \bibinfo {author} {\bibfnamefont {J.~M.}\
  \bibnamefont {Smith}}, \bibinfo {author} {\bibfnamefont {T.~W.}\ \bibnamefont
  {H{\"a}nsch}},\ and\ \bibinfo {author} {\bibfnamefont {D.}~\bibnamefont
  {Hunger}},\ }\bibfield  {title} {\bibinfo {title} {{Cavity-Enhanced
  Single-Photon Source Based on the Silicon-Vacancy Center in Diamond}},\
  }\href {https://link.aps.org/doi/10.1103/PhysRevApplied.7.024031} {\bibfield
  {journal} {\bibinfo  {journal} {Physical Review Applied}\ }\textbf {\bibinfo
  {volume} {7}},\ \bibinfo {pages} {024031} (\bibinfo {year}
  {2017})}\BibitemShut {NoStop}%
\bibitem [{\citenamefont {Hunger}\ \emph {et~al.}(2010)\citenamefont {Hunger},
  \citenamefont {Steinmetz}, \citenamefont {Colombe}, \citenamefont {Deutsch},
  \citenamefont {H{\"a}nsch},\ and\ \citenamefont {Reichel}}]{hunger2010}%
  \BibitemOpen
  \bibfield  {author} {\bibinfo {author} {\bibfnamefont {D.}~\bibnamefont
  {Hunger}}, \bibinfo {author} {\bibfnamefont {T.}~\bibnamefont {Steinmetz}},
  \bibinfo {author} {\bibfnamefont {Y.}~\bibnamefont {Colombe}}, \bibinfo
  {author} {\bibfnamefont {C.}~\bibnamefont {Deutsch}}, \bibinfo {author}
  {\bibfnamefont {T.~W.}\ \bibnamefont {H{\"a}nsch}},\ and\ \bibinfo {author}
  {\bibfnamefont {J.}~\bibnamefont {Reichel}},\ }\bibfield  {title} {\bibinfo
  {title} {{A fiber Fabry–Perot cavity with high finesse}},\ }\href
  {https://doi.org/10.1088/1367-2630/12/6/065038} {\bibfield  {journal}
  {\bibinfo  {journal} {New Journal of Physics}\ }\textbf {\bibinfo {volume}
  {12}},\ \bibinfo {pages} {065038} (\bibinfo {year} {2010})}\BibitemShut
  {NoStop}%
\bibitem [{Note1()}]{Note1}%
  \BibitemOpen
  \bibinfo {note} {\label {suppnote}See Supplemental Material at [URL will be
  inserted by publisher] for further information regarding membrane
  fabrication, experimental characterization, and theoretical details. The
  supplementary includes Refs. [50-62]}\BibitemShut {NoStop}%
\bibitem [{\citenamefont {Chen}\ \emph {et~al.}(2019)\citenamefont {Chen},
  \citenamefont {Mu}, \citenamefont {Zhou}, \citenamefont {Fr\"och},
  \citenamefont {Rasmit}, \citenamefont {Diederichs}, \citenamefont {Zheludev},
  \citenamefont {Aharonovich},\ and\ \citenamefont {Gao}}]{chen2019}%
  \BibitemOpen
  \bibfield  {author} {\bibinfo {author} {\bibfnamefont {D.}~\bibnamefont
  {Chen}}, \bibinfo {author} {\bibfnamefont {Z.}~\bibnamefont {Mu}}, \bibinfo
  {author} {\bibfnamefont {Y.}~\bibnamefont {Zhou}}, \bibinfo {author}
  {\bibfnamefont {J.~E.}\ \bibnamefont {Fr\"och}}, \bibinfo {author}
  {\bibfnamefont {A.}~\bibnamefont {Rasmit}}, \bibinfo {author} {\bibfnamefont
  {C.}~\bibnamefont {Diederichs}}, \bibinfo {author} {\bibfnamefont
  {N.}~\bibnamefont {Zheludev}}, \bibinfo {author} {\bibfnamefont
  {I.}~\bibnamefont {Aharonovich}},\ and\ \bibinfo {author} {\bibfnamefont
  {W.-b.}\ \bibnamefont {Gao}},\ }\bibfield  {title} {\bibinfo {title}
  {{Optical Gating of Resonance Fluorescence from a Single Germanium Vacancy
  Color Center in Diamond}},\ }\href
  {https://link.aps.org/doi/10.1103/PhysRevLett.123.033602} {\bibfield
  {journal} {\bibinfo  {journal} {Physical Revivew Letters}\ }\textbf {\bibinfo
  {volume} {123}},\ \bibinfo {pages} {033602} (\bibinfo {year}
  {2019})}\BibitemShut {NoStop}%
\bibitem [{\citenamefont {Brouri}\ \emph {et~al.}(2000)\citenamefont {Brouri},
  \citenamefont {Beveratos}, \citenamefont {Poizat},\ and\ \citenamefont
  {Grangier}}]{Brouri2000}%
  \BibitemOpen
  \bibfield  {author} {\bibinfo {author} {\bibfnamefont {R.}~\bibnamefont
  {Brouri}}, \bibinfo {author} {\bibfnamefont {A.}~\bibnamefont {Beveratos}},
  \bibinfo {author} {\bibfnamefont {J.-P.}\ \bibnamefont {Poizat}},\ and\
  \bibinfo {author} {\bibfnamefont {P.}~\bibnamefont {Grangier}},\ }\bibfield
  {title} {\bibinfo {title} {Photon antibunching in the fluorescence of
  individual color centers in diamond},\ }\href
  {https://doi.org/10.1364/ol.25.001294} {\bibfield  {journal} {\bibinfo
  {journal} {Optics Letters}\ }\textbf {\bibinfo {volume} {25}},\ \bibinfo
  {pages} {1294} (\bibinfo {year} {2000})}\BibitemShut {NoStop}%
\bibitem [{\citenamefont {Neu}\ \emph {et~al.}(2012)\citenamefont {Neu},
  \citenamefont {Agio},\ and\ \citenamefont {Becher}}]{Neu2012}%
  \BibitemOpen
  \bibfield  {author} {\bibinfo {author} {\bibfnamefont {E.}~\bibnamefont
  {Neu}}, \bibinfo {author} {\bibfnamefont {M.}~\bibnamefont {Agio}},\ and\
  \bibinfo {author} {\bibfnamefont {C.}~\bibnamefont {Becher}},\ }\bibfield
  {title} {\bibinfo {title} {Photophysics of single silicon vacancy centers in
  diamond: implications for single photon emission},\ }\href
  {https://doi.org/10.1364/OE.20.019956} {\bibfield  {journal} {\bibinfo
  {journal} {Optics Express}\ }\textbf {\bibinfo {volume} {20}},\ \bibinfo
  {pages} {19956} (\bibinfo {year} {2012})}\BibitemShut {NoStop}%
\bibitem [{\citenamefont {Boldyrev}\ \emph {et~al.}(2018)\citenamefont
  {Boldyrev}, \citenamefont {Mavrin}, \citenamefont {Sherin},\ and\
  \citenamefont {Popova}}]{boldyrev2018}%
  \BibitemOpen
  \bibfield  {author} {\bibinfo {author} {\bibfnamefont {K.}~\bibnamefont
  {Boldyrev}}, \bibinfo {author} {\bibfnamefont {B.}~\bibnamefont {Mavrin}},
  \bibinfo {author} {\bibfnamefont {P.~S.}\ \bibnamefont {Sherin}},\ and\
  \bibinfo {author} {\bibfnamefont {M.}~\bibnamefont {Popova}},\ }\bibfield
  {title} {\bibinfo {title} {{Bright luminescence of diamonds with Ge-V
  centers}},\ }\href {https://doi.org/10.1016/j.jlumin.2017.07.031} {\bibfield
  {journal} {\bibinfo  {journal} {Journal of Luminescence}\ }\textbf {\bibinfo
  {volume} {193}},\ \bibinfo {pages} {119} (\bibinfo {year}
  {2018})}\BibitemShut {NoStop}%
\bibitem [{\citenamefont {Nguyen}\ \emph
  {et~al.}(2019{\natexlab{b}})\citenamefont {Nguyen}, \citenamefont {Nikolay},
  \citenamefont {Bradac}, \citenamefont {Kianinia}, \citenamefont {Ekimov},
  \citenamefont {Mendelson}, \citenamefont {Benson},\ and\ \citenamefont
  {Aharonovich}}]{Nguyen2019-gev}%
  \BibitemOpen
  \bibfield  {author} {\bibinfo {author} {\bibfnamefont {M.}~\bibnamefont
  {Nguyen}}, \bibinfo {author} {\bibfnamefont {N.}~\bibnamefont {Nikolay}},
  \bibinfo {author} {\bibfnamefont {C.}~\bibnamefont {Bradac}}, \bibinfo
  {author} {\bibfnamefont {M.}~\bibnamefont {Kianinia}}, \bibinfo {author}
  {\bibfnamefont {E.~A.}\ \bibnamefont {Ekimov}}, \bibinfo {author}
  {\bibfnamefont {N.}~\bibnamefont {Mendelson}}, \bibinfo {author}
  {\bibfnamefont {O.}~\bibnamefont {Benson}},\ and\ \bibinfo {author}
  {\bibfnamefont {I.}~\bibnamefont {Aharonovich}},\ }\bibfield  {title}
  {\bibinfo {title} {{Photodynamics and quantum efficiency of germanium vacancy
  color centers in diamond}},\ }\href {https://doi.org/10.1117/1.ap.1.6.066002}
  {\bibfield  {journal} {\bibinfo  {journal} {Advanced Photonics}\ }\textbf
  {\bibinfo {volume} {1}},\ \bibinfo {pages} {1} (\bibinfo {year}
  {2019}{\natexlab{b}})}\BibitemShut {NoStop}%
\bibitem [{\citenamefont {H\"{a}u{\ss}ler}\ \emph {et~al.}(2017)\citenamefont
  {H\"{a}u{\ss}ler}, \citenamefont {Thiering}, \citenamefont {Dietrich},
  \citenamefont {Waasem}, \citenamefont {Teraji}, \citenamefont {Isoya},
  \citenamefont {Iwasaki}, \citenamefont {Hatano}, \citenamefont {Jelezko},
  \citenamefont {Gali} \emph {et~al.}}]{hausler2017}%
  \BibitemOpen
  \bibfield  {author} {\bibinfo {author} {\bibfnamefont {S.}~\bibnamefont
  {H\"{a}u{\ss}ler}}, \bibinfo {author} {\bibfnamefont {G.}~\bibnamefont
  {Thiering}}, \bibinfo {author} {\bibfnamefont {A.}~\bibnamefont {Dietrich}},
  \bibinfo {author} {\bibfnamefont {N.}~\bibnamefont {Waasem}}, \bibinfo
  {author} {\bibfnamefont {T.}~\bibnamefont {Teraji}}, \bibinfo {author}
  {\bibfnamefont {J.}~\bibnamefont {Isoya}}, \bibinfo {author} {\bibfnamefont
  {T.}~\bibnamefont {Iwasaki}}, \bibinfo {author} {\bibfnamefont
  {M.}~\bibnamefont {Hatano}}, \bibinfo {author} {\bibfnamefont
  {F.}~\bibnamefont {Jelezko}}, \bibinfo {author} {\bibfnamefont
  {A.}~\bibnamefont {Gali}}, \emph {et~al.},\ }\bibfield  {title} {\bibinfo
  {title} {Photoluminescence excitation spectroscopy of {SiV$^{-}$} and
  {GeV$^{-}$} color center in diamond},\ }\href
  {https://doi.org/10.1088%2F1367-2630%2Faa73e5} {\bibfield  {journal}
  {\bibinfo  {journal} {New Journal of Physics}\ }\textbf {\bibinfo {volume}
  {19}},\ \bibinfo {pages} {063036} (\bibinfo {year} {2017})}\BibitemShut
  {NoStop}%
\bibitem [{\citenamefont {Palyanov}\ \emph {et~al.}(2015)\citenamefont
  {Palyanov}, \citenamefont {Kupriyanov}, \citenamefont {Borzdov},\ and\
  \citenamefont {Surovtsev}}]{palyanov2015}%
  \BibitemOpen
  \bibfield  {author} {\bibinfo {author} {\bibfnamefont {Y.~N.}\ \bibnamefont
  {Palyanov}}, \bibinfo {author} {\bibfnamefont {I.~N.}\ \bibnamefont
  {Kupriyanov}}, \bibinfo {author} {\bibfnamefont {Y.~M.}\ \bibnamefont
  {Borzdov}},\ and\ \bibinfo {author} {\bibfnamefont {N.~V.}\ \bibnamefont
  {Surovtsev}},\ }\bibfield  {title} {\bibinfo {title} {Germanium: a new
  catalyst for diamond synthesis and a new optically active impurity in
  diamond},\ }\href {https://doi.org/10.1038/srep14789} {\bibfield  {journal}
  {\bibinfo  {journal} {Scientific Reports}\ }\textbf {\bibinfo {volume} {5}},\
  \bibinfo {pages} {14789} (\bibinfo {year} {2015})}\BibitemShut {NoStop}%
\bibitem [{\citenamefont {Janitz}\ \emph {et~al.}(2017)\citenamefont {Janitz},
  \citenamefont {Ruf}, \citenamefont {Fontana}, \citenamefont {Sankey},\ and\
  \citenamefont {Childress}}]{janitz2017}%
  \BibitemOpen
  \bibfield  {author} {\bibinfo {author} {\bibfnamefont {E.}~\bibnamefont
  {Janitz}}, \bibinfo {author} {\bibfnamefont {M.}~\bibnamefont {Ruf}},
  \bibinfo {author} {\bibfnamefont {Y.}~\bibnamefont {Fontana}}, \bibinfo
  {author} {\bibfnamefont {J.}~\bibnamefont {Sankey}},\ and\ \bibinfo {author}
  {\bibfnamefont {L.}~\bibnamefont {Childress}},\ }\bibfield  {title} {\bibinfo
  {title} {{High mechanical bandwidth fiber-coupled Fabry-Perot cavity}},\
  }\href {https://doi.org/10.1364/OE.25.020932} {\bibfield  {journal} {\bibinfo
   {journal} {Optics Express}\ }\textbf {\bibinfo {volume} {25}},\ \bibinfo
  {pages} {20932} (\bibinfo {year} {2017})}\BibitemShut {NoStop}%
\bibitem [{\citenamefont {Najer}\ \emph {et~al.}(2017)\citenamefont {Najer},
  \citenamefont {Renggli}, \citenamefont {Riedel}, \citenamefont
  {Starosielec},\ and\ \citenamefont {Warburton}}]{najer2017}%
  \BibitemOpen
  \bibfield  {author} {\bibinfo {author} {\bibfnamefont {D.}~\bibnamefont
  {Najer}}, \bibinfo {author} {\bibfnamefont {M.}~\bibnamefont {Renggli}},
  \bibinfo {author} {\bibfnamefont {D.}~\bibnamefont {Riedel}}, \bibinfo
  {author} {\bibfnamefont {S.}~\bibnamefont {Starosielec}},\ and\ \bibinfo
  {author} {\bibfnamefont {R.~J.}\ \bibnamefont {Warburton}},\ }\bibfield
  {title} {\bibinfo {title} {{Fabrication of mirror templates in silica with
  micron-sized radii of curvature}},\ }\href
  {https://doi.org/10.1063/1.4973458} {\bibfield  {journal} {\bibinfo
  {journal} {Applied Physics Letters}\ }\textbf {\bibinfo {volume} {110}},\
  \bibinfo {pages} {011101} (\bibinfo {year} {2017})}\BibitemShut {NoStop}%
\bibitem [{\citenamefont {Trusheim}\ \emph {et~al.}(2020)\citenamefont
  {Trusheim}, \citenamefont {Pingault}, \citenamefont {Wan}, \citenamefont
  {De~Santis}, \citenamefont {Chen}, \citenamefont {Walsh}, \citenamefont
  {Rose}, \citenamefont {Becker}, \citenamefont {Bersin}, \citenamefont
  {Malladi} \emph {et~al.}}]{trusheim2018}%
  \BibitemOpen
  \bibfield  {author} {\bibinfo {author} {\bibfnamefont {M.~E.}\ \bibnamefont
  {Trusheim}}, \bibinfo {author} {\bibfnamefont {B.}~\bibnamefont {Pingault}},
  \bibinfo {author} {\bibfnamefont {N.~H.}\ \bibnamefont {Wan}}, \bibinfo
  {author} {\bibfnamefont {L.}~\bibnamefont {De~Santis}}, \bibinfo {author}
  {\bibfnamefont {K.~C.}\ \bibnamefont {Chen}}, \bibinfo {author}
  {\bibfnamefont {M.}~\bibnamefont {Walsh}}, \bibinfo {author} {\bibfnamefont
  {J.~J.}\ \bibnamefont {Rose}}, \bibinfo {author} {\bibfnamefont {J.~N.}\
  \bibnamefont {Becker}}, \bibinfo {author} {\bibfnamefont {E.}~\bibnamefont
  {Bersin}}, \bibinfo {author} {\bibfnamefont {G.}~\bibnamefont {Malladi}},
  \emph {et~al.},\ }\bibfield  {title} {\bibinfo {title} {{Transform-Limited
  Photons From a Coherent Tin-Vacancy Spin in Diamond}},\ }\href
  {https://doi.org/10.1103/PhysRevLett.124.023602} {\bibfield  {journal}
  {\bibinfo  {journal} {Physical Review Letters}\ }\textbf {\bibinfo {volume}
  {124}},\ \bibinfo {pages} {023602} (\bibinfo {year} {2020})}\BibitemShut
  {NoStop}%
\bibitem [{\citenamefont {Rugar}\ \emph {et~al.}(2019)\citenamefont {Rugar},
  \citenamefont {Dory}, \citenamefont {Sun},\ and\ \citenamefont
  {Vu{\v{c}}kovi{\'c}}}]{rugar2019}%
  \BibitemOpen
  \bibfield  {author} {\bibinfo {author} {\bibfnamefont {A.~E.}\ \bibnamefont
  {Rugar}}, \bibinfo {author} {\bibfnamefont {C.}~\bibnamefont {Dory}},
  \bibinfo {author} {\bibfnamefont {S.}~\bibnamefont {Sun}},\ and\ \bibinfo
  {author} {\bibfnamefont {J.}~\bibnamefont {Vu{\v{c}}kovi{\'c}}},\ }\bibfield
  {title} {\bibinfo {title} {Characterization of optical and spin properties of
  single tin-vacancy centers in diamond nanopillars},\ }\href
  {https://link.aps.org/doi/10.1103/PhysRevB.99.205417} {\bibfield  {journal}
  {\bibinfo  {journal} {Physical Review B}\ }\textbf {\bibinfo {volume} {99}},\
  \bibinfo {pages} {205417} (\bibinfo {year} {2019})}\BibitemShut {NoStop}%
\bibitem [{\citenamefont {Trusheim}\ \emph {et~al.}(2019)\citenamefont
  {Trusheim}, \citenamefont {Wan}, \citenamefont {Chen}, \citenamefont
  {Ciccarino}, \citenamefont {Flick}, \citenamefont {Sundararaman},
  \citenamefont {Malladi}, \citenamefont {Bersin}, \citenamefont {Walsh},
  \citenamefont {Lienhard} \emph {et~al.}}]{trusheim2019}%
  \BibitemOpen
  \bibfield  {author} {\bibinfo {author} {\bibfnamefont {M.~E.}\ \bibnamefont
  {Trusheim}}, \bibinfo {author} {\bibfnamefont {N.~H.}\ \bibnamefont {Wan}},
  \bibinfo {author} {\bibfnamefont {K.~C.}\ \bibnamefont {Chen}}, \bibinfo
  {author} {\bibfnamefont {C.~J.}\ \bibnamefont {Ciccarino}}, \bibinfo {author}
  {\bibfnamefont {J.}~\bibnamefont {Flick}}, \bibinfo {author} {\bibfnamefont
  {R.}~\bibnamefont {Sundararaman}}, \bibinfo {author} {\bibfnamefont
  {G.}~\bibnamefont {Malladi}}, \bibinfo {author} {\bibfnamefont
  {E.}~\bibnamefont {Bersin}}, \bibinfo {author} {\bibfnamefont
  {M.}~\bibnamefont {Walsh}}, \bibinfo {author} {\bibfnamefont
  {B.}~\bibnamefont {Lienhard}}, \emph {et~al.},\ }\bibfield  {title} {\bibinfo
  {title} {Lead-related quantum emitters in diamond},\ }\href@noop {}
  {\bibfield  {journal} {\bibinfo  {journal} {Physical Review B}\ }\textbf
  {\bibinfo {volume} {99}},\ \bibinfo {pages} {075430} (\bibinfo {year}
  {2019})}\BibitemShut {NoStop}%
\bibitem [{\citenamefont {Tchernij}\ \emph {et~al.}(2018)\citenamefont
  {Tchernij}, \citenamefont {L\"{u}hmann}, \citenamefont {Herzig},
  \citenamefont {K\"{u}pper}, \citenamefont {Damin}, \citenamefont
  {Santonocito}, \citenamefont {Signorile}, \citenamefont {Traina},
  \citenamefont {Moreva}, \citenamefont {Celegato} \emph
  {et~al.}}]{DitaliaTchernij2018}%
  \BibitemOpen
  \bibfield  {author} {\bibinfo {author} {\bibfnamefont {S.~D.}\ \bibnamefont
  {Tchernij}}, \bibinfo {author} {\bibfnamefont {T.}~\bibnamefont
  {L\"{u}hmann}}, \bibinfo {author} {\bibfnamefont {T.}~\bibnamefont {Herzig}},
  \bibinfo {author} {\bibfnamefont {J.}~\bibnamefont {K\"{u}pper}}, \bibinfo
  {author} {\bibfnamefont {A.}~\bibnamefont {Damin}}, \bibinfo {author}
  {\bibfnamefont {S.}~\bibnamefont {Santonocito}}, \bibinfo {author}
  {\bibfnamefont {M.}~\bibnamefont {Signorile}}, \bibinfo {author}
  {\bibfnamefont {P.}~\bibnamefont {Traina}}, \bibinfo {author} {\bibfnamefont
  {E.}~\bibnamefont {Moreva}}, \bibinfo {author} {\bibfnamefont
  {F.}~\bibnamefont {Celegato}}, \emph {et~al.},\ }\bibfield  {title} {\bibinfo
  {title} {{Single-Photon Emitters in Lead-Implanted Single-Crystal Diamond}},\
  }\href {https://doi.org/10.1021/acsphotonics.8b01013} {\bibfield  {journal}
  {\bibinfo  {journal} {{ACS} Photonics}\ }\textbf {\bibinfo {volume} {5}},\
  \bibinfo {pages} {4864} (\bibinfo {year} {2018})}\BibitemShut {NoStop}%
\bibitem [{\citenamefont {Hausmann}\ \emph {et~al.}(2012)\citenamefont
  {Hausmann}, \citenamefont {Shields}, \citenamefont {Quan}, \citenamefont
  {Maletinsky}, \citenamefont {McCutcheon}, \citenamefont {Choy}, \citenamefont
  {Babinec}, \citenamefont {Kubanek}, \citenamefont {Yacoby}, \citenamefont
  {Lukin} \emph {et~al.}}]{hausmann2012}%
  \BibitemOpen
  \bibfield  {author} {\bibinfo {author} {\bibfnamefont {B.~J.~M.}\
  \bibnamefont {Hausmann}}, \bibinfo {author} {\bibfnamefont {B.}~\bibnamefont
  {Shields}}, \bibinfo {author} {\bibfnamefont {Q.}~\bibnamefont {Quan}},
  \bibinfo {author} {\bibfnamefont {P.}~\bibnamefont {Maletinsky}}, \bibinfo
  {author} {\bibfnamefont {M.}~\bibnamefont {McCutcheon}}, \bibinfo {author}
  {\bibfnamefont {J.~T.}\ \bibnamefont {Choy}}, \bibinfo {author}
  {\bibfnamefont {T.~M.}\ \bibnamefont {Babinec}}, \bibinfo {author}
  {\bibfnamefont {A.}~\bibnamefont {Kubanek}}, \bibinfo {author} {\bibfnamefont
  {A.}~\bibnamefont {Yacoby}}, \bibinfo {author} {\bibfnamefont {M.~D.}\
  \bibnamefont {Lukin}}, \emph {et~al.},\ }\bibfield  {title} {\bibinfo {title}
  {{Integrated Diamond Networks for Quantum Nanophotonics}},\ }\href
  {https://doi.org/10.1021/nl204449n} {\bibfield  {journal} {\bibinfo
  {journal} {Nano Letters}\ }\textbf {\bibinfo {volume} {12}},\ \bibinfo
  {pages} {1578} (\bibinfo {year} {2012})}\BibitemShut {NoStop}%
\bibitem [{\citenamefont {Latawiec}\ \emph {et~al.}(2015)\citenamefont
  {Latawiec}, \citenamefont {Venkataraman}, \citenamefont {Burek},
  \citenamefont {Hausmann}, \citenamefont {Bulu},\ and\ \citenamefont
  {Lon{\v{c}}ar}}]{latawiec2015}%
  \BibitemOpen
  \bibfield  {author} {\bibinfo {author} {\bibfnamefont {P.}~\bibnamefont
  {Latawiec}}, \bibinfo {author} {\bibfnamefont {V.}~\bibnamefont
  {Venkataraman}}, \bibinfo {author} {\bibfnamefont {M.~J.}\ \bibnamefont
  {Burek}}, \bibinfo {author} {\bibfnamefont {B.~J.}\ \bibnamefont {Hausmann}},
  \bibinfo {author} {\bibfnamefont {I.}~\bibnamefont {Bulu}},\ and\ \bibinfo
  {author} {\bibfnamefont {M.}~\bibnamefont {Lon{\v{c}}ar}},\ }\bibfield
  {title} {\bibinfo {title} {{On-chip diamond Raman laser}},\ }\href
  {https://doi.org/10.1364/OPTICA.2.000924} {\bibfield  {journal} {\bibinfo
  {journal} {Optica}\ }\textbf {\bibinfo {volume} {2}},\ \bibinfo {pages} {924}
  (\bibinfo {year} {2015})}\BibitemShut {NoStop}%
\bibitem [{\citenamefont {Appel}\ \emph {et~al.}(2016)\citenamefont {Appel},
  \citenamefont {Neu}, \citenamefont {Ganzhorn}, \citenamefont {Barfuss},
  \citenamefont {Batzer}, \citenamefont {Gratz}, \citenamefont {Tsch{\"o}pe},\
  and\ \citenamefont {Maletinsky}}]{appel2016}%
  \BibitemOpen
  \bibfield  {author} {\bibinfo {author} {\bibfnamefont {P.}~\bibnamefont
  {Appel}}, \bibinfo {author} {\bibfnamefont {E.}~\bibnamefont {Neu}}, \bibinfo
  {author} {\bibfnamefont {M.}~\bibnamefont {Ganzhorn}}, \bibinfo {author}
  {\bibfnamefont {A.}~\bibnamefont {Barfuss}}, \bibinfo {author} {\bibfnamefont
  {M.}~\bibnamefont {Batzer}}, \bibinfo {author} {\bibfnamefont
  {M.}~\bibnamefont {Gratz}}, \bibinfo {author} {\bibfnamefont
  {A.}~\bibnamefont {Tsch{\"o}pe}},\ and\ \bibinfo {author} {\bibfnamefont
  {P.}~\bibnamefont {Maletinsky}},\ }\bibfield  {title} {\bibinfo {title}
  {Fabrication of all diamond scanning probes for nanoscale magnetometry},\
  }\href {https://doi.org/10.1063/1.4952953} {\bibfield  {journal} {\bibinfo
  {journal} {Review of Scientific Instruments}\ }\textbf {\bibinfo {volume}
  {87}},\ \bibinfo {pages} {063703} (\bibinfo {year} {2016})}\BibitemShut
  {NoStop}%
\bibitem [{\citenamefont {Ziegler}\ \emph {et~al.}(2010)\citenamefont
  {Ziegler}, \citenamefont {Ziegler},\ and\ \citenamefont {Biersack}}]{srim}%
  \BibitemOpen
  \bibfield  {author} {\bibinfo {author} {\bibfnamefont {J.~F.}\ \bibnamefont
  {Ziegler}}, \bibinfo {author} {\bibfnamefont {M.~D.}\ \bibnamefont
  {Ziegler}},\ and\ \bibinfo {author} {\bibfnamefont {J.~P.}\ \bibnamefont
  {Biersack}},\ }\bibfield  {title} {\bibinfo {title} {{SRIM --The stopping and
  range of ions in matter (2010)}},\ }\href
  {http://www.sciencedirect.com/science/article/pii/S0168583X10001862}
  {\bibfield  {journal} {\bibinfo  {journal} {Nuclear Instruments and Methods
  in Physics Research Section B: Beam Interactions with Materials and Atoms}\
  }\textbf {\bibinfo {volume} {268}},\ \bibinfo {pages} {1818} (\bibinfo {year}
  {2010})}\BibitemShut {NoStop}%
\bibitem [{\citenamefont {Chu}\ \emph {et~al.}(2014)\citenamefont {Chu},
  \citenamefont {de~Leon}, \citenamefont {Shields}, \citenamefont {Hausmann},
  \citenamefont {Evans}, \citenamefont {Togan}, \citenamefont {Burek},
  \citenamefont {Markham}, \citenamefont {Stacey}, \citenamefont {Zibrov} \emph
  {et~al.}}]{chu2014}%
  \BibitemOpen
  \bibfield  {author} {\bibinfo {author} {\bibfnamefont {Y.}~\bibnamefont
  {Chu}}, \bibinfo {author} {\bibfnamefont {N.}~\bibnamefont {de~Leon}},
  \bibinfo {author} {\bibfnamefont {B.}~\bibnamefont {Shields}}, \bibinfo
  {author} {\bibfnamefont {B.}~\bibnamefont {Hausmann}}, \bibinfo {author}
  {\bibfnamefont {R.}~\bibnamefont {Evans}}, \bibinfo {author} {\bibfnamefont
  {E.}~\bibnamefont {Togan}}, \bibinfo {author} {\bibfnamefont {M.~J.}\
  \bibnamefont {Burek}}, \bibinfo {author} {\bibfnamefont {M.}~\bibnamefont
  {Markham}}, \bibinfo {author} {\bibfnamefont {A.}~\bibnamefont {Stacey}},
  \bibinfo {author} {\bibfnamefont {A.}~\bibnamefont {Zibrov}}, \emph
  {et~al.},\ }\bibfield  {title} {\bibinfo {title} {{Coherent Optical
  Transitions in Implanted Nitrogen Vacancy Centers}},\ }\href
  {https://doi.org/10.1021/nl404836p} {\bibfield  {journal} {\bibinfo
  {journal} {Nano Letters}\ }\textbf {\bibinfo {volume} {14}},\ \bibinfo
  {pages} {1982} (\bibinfo {year} {2014})}\BibitemShut {NoStop}%
\bibitem [{\citenamefont {Lukosz}\ and\ \citenamefont
  {Kunz}(1977)}]{Lukosz1977}%
  \BibitemOpen
  \bibfield  {author} {\bibinfo {author} {\bibfnamefont {W.}~\bibnamefont
  {Lukosz}}\ and\ \bibinfo {author} {\bibfnamefont {R.~E.}\ \bibnamefont
  {Kunz}},\ }\bibfield  {title} {\bibinfo {title} {{Light emission by magnetic
  and electric dipoles close to a plane interface. I. Total radiated power}},\
  }\href {https://doi.org/10.1364/JOSA.67.001607} {\bibfield  {journal}
  {\bibinfo  {journal} {Journal of the Optical Society of America}\ }\textbf
  {\bibinfo {volume} {67}},\ \bibinfo {pages} {1607} (\bibinfo {year}
  {1977})}\BibitemShut {NoStop}%
\bibitem [{\citenamefont {Neyts}(1998)}]{Neyts1998}%
  \BibitemOpen
  \bibfield  {author} {\bibinfo {author} {\bibfnamefont {K.~A.}\ \bibnamefont
  {Neyts}},\ }\bibfield  {title} {\bibinfo {title} {Simulation of light
  emission from thin-film microcavities},\ }\href
  {https://doi.org/10.1364/JOSAA.15.000962} {\bibfield  {journal} {\bibinfo
  {journal} {Journal of the Optical Society of America A}\ }\textbf {\bibinfo
  {volume} {15}},\ \bibinfo {pages} {962} (\bibinfo {year} {1998})}\BibitemShut
  {NoStop}%
\bibitem [{\citenamefont {Polereck{\`y}}\ \emph {et~al.}(2000)\citenamefont
  {Polereck{\`y}}, \citenamefont {Hamrle},\ and\ \citenamefont
  {MacCraith}}]{polerecky2000}%
  \BibitemOpen
  \bibfield  {author} {\bibinfo {author} {\bibfnamefont {L.}~\bibnamefont
  {Polereck{\`y}}}, \bibinfo {author} {\bibfnamefont {J.}~\bibnamefont
  {Hamrle}},\ and\ \bibinfo {author} {\bibfnamefont {B.~D.}\ \bibnamefont
  {MacCraith}},\ }\bibfield  {title} {\bibinfo {title} {Theory of the radiation
  of dipoles placed within a multilayer system},\ }\href
  {https://doi.org/10.1364/AO.39.003968} {\bibfield  {journal} {\bibinfo
  {journal} {Applied Optics}\ }\textbf {\bibinfo {volume} {39}},\ \bibinfo
  {pages} {3968} (\bibinfo {year} {2000})}\BibitemShut {NoStop}%
\bibitem [{\citenamefont {Reed}\ \emph {et~al.}(1987)\citenamefont {Reed},
  \citenamefont {Giergiel}, \citenamefont {Hemminger},\ and\ \citenamefont
  {Ushioda}}]{reed1987}%
  \BibitemOpen
  \bibfield  {author} {\bibinfo {author} {\bibfnamefont {C.}~\bibnamefont
  {Reed}}, \bibinfo {author} {\bibfnamefont {J.}~\bibnamefont {Giergiel}},
  \bibinfo {author} {\bibfnamefont {J.}~\bibnamefont {Hemminger}},\ and\
  \bibinfo {author} {\bibfnamefont {S.}~\bibnamefont {Ushioda}},\ }\bibfield
  {title} {\bibinfo {title} {Dipole radiation in a multilayer geometry},\
  }\href {https://link.aps.org/doi/10.1103/PhysRevB.36.4990} {\bibfield
  {journal} {\bibinfo  {journal} {Physical Review B}\ }\textbf {\bibinfo
  {volume} {36}},\ \bibinfo {pages} {4990} (\bibinfo {year}
  {1987})}\BibitemShut {NoStop}%
\bibitem [{\citenamefont {Katsidis}\ and\ \citenamefont
  {Siapkas}(2002)}]{katsidis2002}%
  \BibitemOpen
  \bibfield  {author} {\bibinfo {author} {\bibfnamefont {C.~C.}\ \bibnamefont
  {Katsidis}}\ and\ \bibinfo {author} {\bibfnamefont {D.~I.}\ \bibnamefont
  {Siapkas}},\ }\bibfield  {title} {\bibinfo {title} {General transfer-matrix
  method for optical multilayer systems with coherent, partially coherent, and
  incoherent interference},\ }\href {https://doi.org/10.1364/AO.41.003978}
  {\bibfield  {journal} {\bibinfo  {journal} {Applied Optics}\ }\textbf
  {\bibinfo {volume} {41}},\ \bibinfo {pages} {3978} (\bibinfo {year}
  {2002})}\BibitemShut {NoStop}%
\bibitem [{\citenamefont {Arnon}(1977)}]{arnon1977}%
  \BibitemOpen
  \bibfield  {author} {\bibinfo {author} {\bibfnamefont {O.}~\bibnamefont
  {Arnon}},\ }\bibfield  {title} {\bibinfo {title} {Loss mechanisms in
  dielectric optical interference devices},\ }\href
  {https://doi.org/10.1364/AO.16.002147} {\bibfield  {journal} {\bibinfo
  {journal} {Applied Optics}\ }\textbf {\bibinfo {volume} {16}},\ \bibinfo
  {pages} {2147} (\bibinfo {year} {1977})}\BibitemShut {NoStop}%
\bibitem [{\citenamefont {van Dam}\ \emph {et~al.}(2018)\citenamefont {van
  Dam}, \citenamefont {Ruf},\ and\ \citenamefont {Hanson}}]{vandam2018}%
  \BibitemOpen
  \bibfield  {author} {\bibinfo {author} {\bibfnamefont {S.~B.}\ \bibnamefont
  {van Dam}}, \bibinfo {author} {\bibfnamefont {M.}~\bibnamefont {Ruf}},\ and\
  \bibinfo {author} {\bibfnamefont {R.}~\bibnamefont {Hanson}},\ }\bibfield
  {title} {\bibinfo {title} {Optimal design of diamond-air microcavities for
  quantum networks using an analytical approach},\ }\href
  {http://stacks.iop.org/1367-2630/20/i=11/a=115004} {\bibfield  {journal}
  {\bibinfo  {journal} {New Journal of Physics}\ }\textbf {\bibinfo {volume}
  {20}},\ \bibinfo {pages} {115004} (\bibinfo {year} {2018})}\BibitemShut
  {NoStop}%
\bibitem [{\citenamefont {Solin}\ and\ \citenamefont
  {Ramdas}(1970)}]{solin1970}%
  \BibitemOpen
  \bibfield  {author} {\bibinfo {author} {\bibfnamefont {S.~A.}\ \bibnamefont
  {Solin}}\ and\ \bibinfo {author} {\bibfnamefont {A.~K.}\ \bibnamefont
  {Ramdas}},\ }\bibfield  {title} {\bibinfo {title} {{Raman Spectrum of
  Diamond}},\ }\href {https://link.aps.org/doi/10.1103/PhysRevB.1.1687}
  {\bibfield  {journal} {\bibinfo  {journal} {Physical Review B}\ }\textbf
  {\bibinfo {volume} {1}},\ \bibinfo {pages} {1687} (\bibinfo {year}
  {1970})}\BibitemShut {NoStop}%
\end{thebibliography}%


%

\end{document}